\documentclass[acmtog]{acmart}

\usepackage{booktabs} 

\usepackage{graphicx}
\usepackage{subfigure}
\usepackage{amssymb}
\usepackage{amsmath}
\usepackage{url}
\usepackage{color}
\usepackage{wrapfig}
\usepackage{dblfloatfix}
\usepackage{enumitem}
\usepackage{microtype}

\usepackage[ruled]{algorithm2e} 

\SetAlFnt{\small}
\SetAlCapFnt{\small}
\SetAlCapNameFnt{\small}
\SetAlCapHSkip{0pt}

\setcopyright{acmcopyright}
\acmJournal{TOG}
\acmYear{2019}\acmVolume{38}\acmNumber{6}\acmArticle{196}\acmMonth{11} \acmDOI{10.1145/3355089.3356504}

\begin{document}

\newcommand{\TODO}[1]{{\color{red}{[TODO: #1]}}}
\newcommand{\rz}[1]{{\color{black}#1}}
\newcommand{\rzf}[1]{{\color{black}#1}}
\newcommand{\xh}[1]{{\color{black}#1}}
\newcommand{\ed}[1]{{\color{black}#1}}
\newcommand{\new}[1]{{\color{black}#1}}
\newcommand{\phil}[1]{{\color{black}#1}}
\newcommand{\final}[1]{{\color{black}#1}}

\newcommand{\legomark}{LEGO\textsuperscript{{\tiny \textregistered}}}
\newcommand{\legoclaim}{}

\newcommand{\pos}{\mathbf{p}}
\newcommand{\posmat}{\mathbf{P}}

\title{Computational LEGO\textsuperscript{{\footnotesize \textregistered}} Technic Design}

\begin{abstract}
We introduce a method to automatically compute \legomark\footnote[1]{\legomark is a trademark of the \legomark Group, which does not sponsor, authorize or endorse this work. All information in this paper is collected and interpreted by its authors and does not represent the opinion of the LEGO Group.  Permission (images in paper) granted by \legomark. All rights reserved.\vspace{-1.0mm}} Technic models from user input sketches, \rz{optionally with motion annotations.}
The generated models resemble the input sketches with coherently-connected bricks and simple layouts, while respecting the intended symmetry \rz{and mechanical properties expressed in the inputs.}
This complex computational assembly problem involves an immense search space, and a much richer brick set and connection mechanisms than regular \legomark.
To address it, we first comprehensively model the brick properties and connection mechanisms, then formulate the construction requirements into an objective function, accounting for faithfulness to input sketch, model simplicity, and structural integrity.
\rz{Next, we model the problem as a sketch cover, where we iteratively refine a random initial layout to cover the input sketch, while guided by the objective.}
At last, we provide a working system to analyze the balance, stress, and assemblability of the generated model.
\rz{To evaluate our method, we compared it with four baselines and professional designs by a \legomark expert, demonstrating the superiority of our automatic designs.}
Also, we recruited several users to try our system, employed it to create models of varying forms and complexities, and physically built most of them.
\end{abstract}

\author{Hao Xu}
\author{Ka-Hei Hui}
\author{Chi-Wing Fu}
\affiliation{%
	\institution{The Chinese University of Hong Kong}}
\author{Hao Zhang}
\affiliation{%
	\institution{Simon Fraser University}}
\renewcommand\shortauthors{Xu. et al.}

\ccsdesc[500]{Computing methodologies~Shape modeling}


\begin{teaserfigure}
	\centering
	\vspace*{-2mm}
	\includegraphics[width=16.8cm]{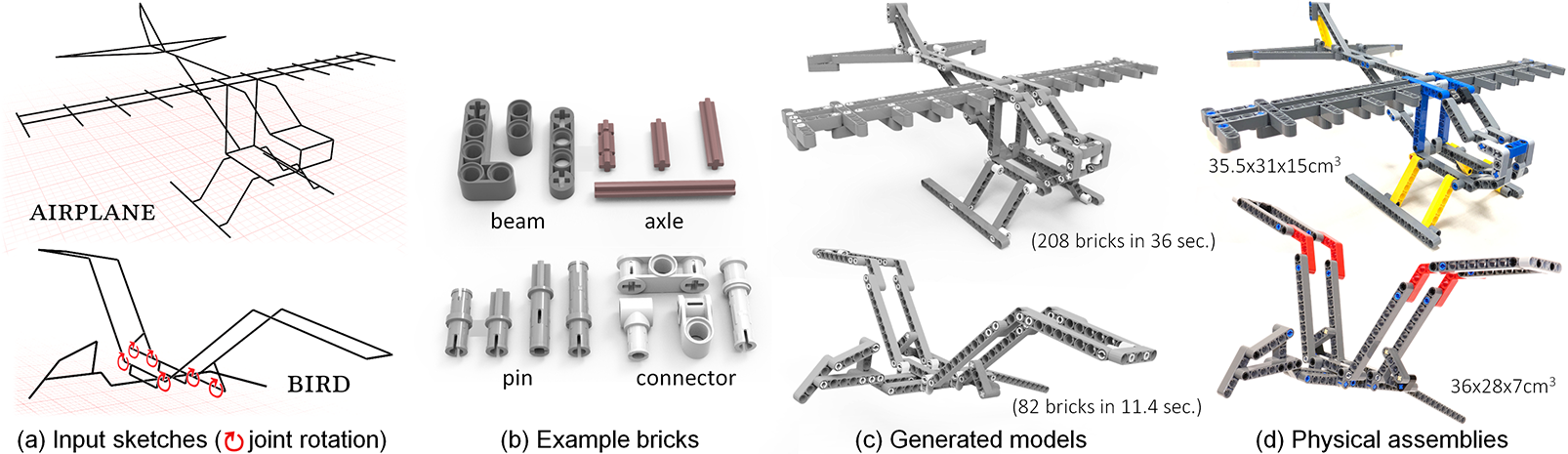}
	\vspace*{-2mm}
	\caption{We introduce a fully automatic method to compute \legomark Technic designs (c \& d) that resemble user input sketches, \rz{which may be annotated with joint rotations (bottom-left).}
	The designs are built by coherently-connected \legomark Technic bricks (b) that respect the symmetry and structural integrity of the models, as well as the \phil{dynamics implied at the specified joints}, while aiming for minimalistic layouts. \legoclaim}
%
	\label{fig:teaser}
	\vspace*{1mm}
\end{teaserfigure}

\maketitle


\section{Introduction}
\label{sec:introduction}

\begin{figure}[!t]
  \centering
  \includegraphics[width=0.99\linewidth]{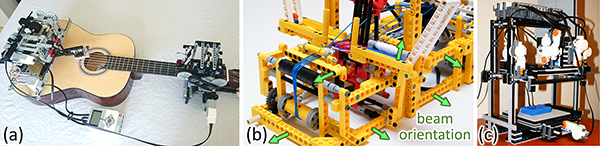}
  \vspace*{-1.5mm}
  \caption{\legomark Technic models designed by enthusiasts:
(a) a guitar robot by YouTuber TECHNICally Possible (see~``\url{https://www.youtube.com/watch?v=cXgB3lIvPHI}'' with over \new{``20.6M''\/} views);
(b) a mechanical loom by N. Lespour (see~``\url{https://www.youtube.com/watch?v=TKdUPbtE_xk}''); 
and (c) a 3D \legomark printer coined MakerLegoBot by W. Gorman (see~``\url{http://www.battlebricks.com/makerlegobot/}''). \legoclaim}
  \label{fig:examples}
  \vspace*{-1mm}
\end{figure}

The \legomark Technic system~\cite{Lego-2018-Technic} was introduced as an expert series for building advanced 3D models in 1977.
%
With brick pieces well\/ beyond those in regular \legomark, e.g., beams, axles, pins, and connectors, as shown in Figure~\ref{fig:teaser}(b), 
%
one can build 3D frame-based structures like those commonly-seen in architecture, \rz{as well as
mechanical assemblies \phil{that} exhibit dynamic behavior such as joint rotations.}
\legomark Technic designs have resulted in a variety of customizable robotics and mechanical models; see Figure~\ref{fig:examples} for some elaborative designs by enthusiasts.
%


Designing \legomark Technic models is considerably more challenging than regular \legomark models, even without adding complex mechanical elements such as gears and pulleys.
Compared with regular \legomark bricks, which are \final{mostly} connected through studs on top of the bricks, \legomark Technic bricks are connected in a variety of ways, also via different kinds of pins and connectors, some of which \rz{allow joint rotations, as shown in Figures~\ref{fig:teaser}(b) and~\ref{fig:brickset_and_connections}.}
%
The sheer number of
assembly varieties leads to an immense search space. For instance, there are over six billion ways of assembling a simple square with nine-unit long sides. Also, as a result of the connection mechanisms, \legomark 
Technic models are often built 
with stricter and more intricate assembly order than regular \legomark 
models. Yet, the Technic system can better adapt to form non-blocky, frame-based, \phil{and even articulable\/} shapes in 3D, since the \legomark Technic beams can be arranged flexibly in different orientations for building 
different parts in the models; see the green arrows in Figure~\ref{fig:examples}(b) that indicate the orientations of the associated beams.


\begin{figure}[!t]
  \centering
  \includegraphics[width=7.8cm]{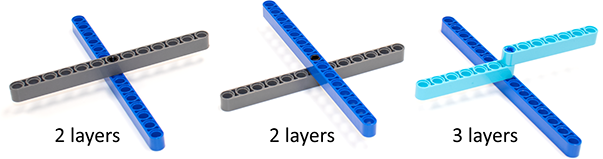}
  \vspace*{-2mm}
  \caption{A cross can be built with different beam layering options. \legoclaim}
  \label{fig:layering}
  \vspace*{-2mm}
\end{figure}

%
Due to the connection mechanisms using pins and connectors, Technic models have an entirely different and far more 
complex building style compared with the simple bottom-up style of regular \legomark brick assemblies.
%
\legomark Technic builders have to mindfully plan the {\em beam placements\/}, {\em orientations\/}, {\em connections\/}, and {\em layering\/} (see Figure~\ref{fig:layering}).
Typically, they need to think several steps ahead and create small assemblies to test the feasibility via \new{trial-and-error\/}~\cite{kmiec-2016-builder-guide,Lego-2018-Technic}.
Given such complexity, manual designs of Technic models are tedious and challenging, requiring substantial expertise and time 
to realize a working design.


In this work, we aim to develop a {\em fully automatic\/} computational method for \legomark Technic model construction. Specifically, given a user-drawn line sketch of a frame-based 3D model, \rz{optionally with annotations of joint rotations, e.g., see Figure~\ref{fig:teaser}(a),} our method
%
{\em automatically selects and arranges \legomark Technic bricks to form Technic assemblies that resemble \phil{and cover\/} the input sketches, while respecting the structural integrity and joint motions of the designs and striving for simplicity and cost-effectiveness of the assemblies.}
%


The computational challenges of this problem stem from the immense search space and the multitude of 
different 
connection mechanisms, beam orientations, and layering options. A viable construction often involves much more than aligning the Technic pieces with the sketched lines. \new{To achieve a coherent, structurally plausible, and functional assembly\/}, the final \legomark Technic models may deviate in nontrivial ways from the input sketches, e.g., see the computed assembly at the tail of the {\sc airplane} in Figure~\ref{fig:teaser}.

%

\begin{figure*}[!t]
  \centering
  \includegraphics[width=16.8cm]{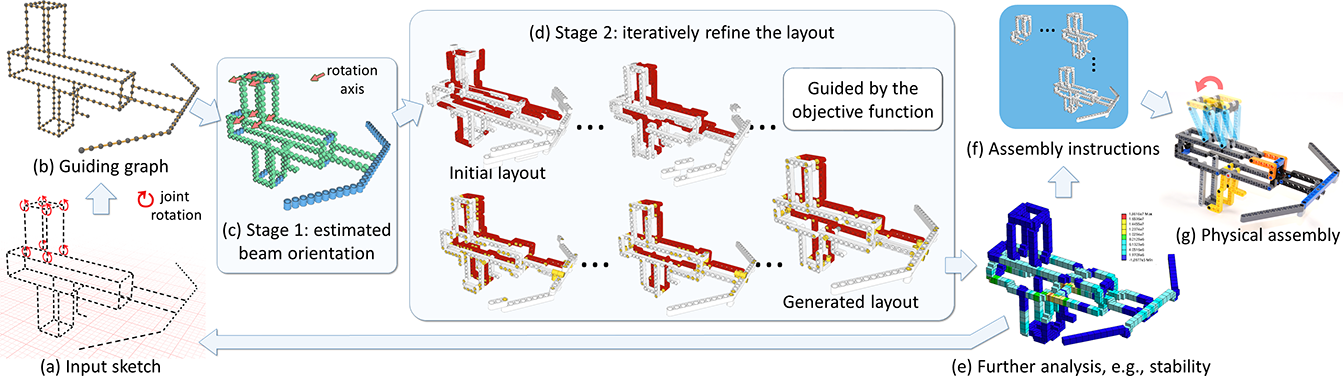}
  \vspace*{-1mm}
  \caption{Overview of our approach.
\phil{First, we construct a guiding graph (b) to abstract the user-input sketch (a), which has optional annotations for joint rotations.
Then, we estimate the local orientation of beams (c), and iteratively refine a random initial layout to cover the input sketch, guided by the objective function and constrained by the specified joint rotations (d).
Further, we analyze the stability, balance, and assemblability of the generated model (e), and produce assembly instructions (f) to facilitate physical construction (g).
Upon the analysis, we may go back and revise the input sketch: (e) $\rightarrow$ (a).} \legoclaim}
%
  \label{fig:overview}
  \vspace*{-1mm}
\end{figure*}
\begin{figure*}[!t]
  \centering
  \includegraphics[height=4.9cm]{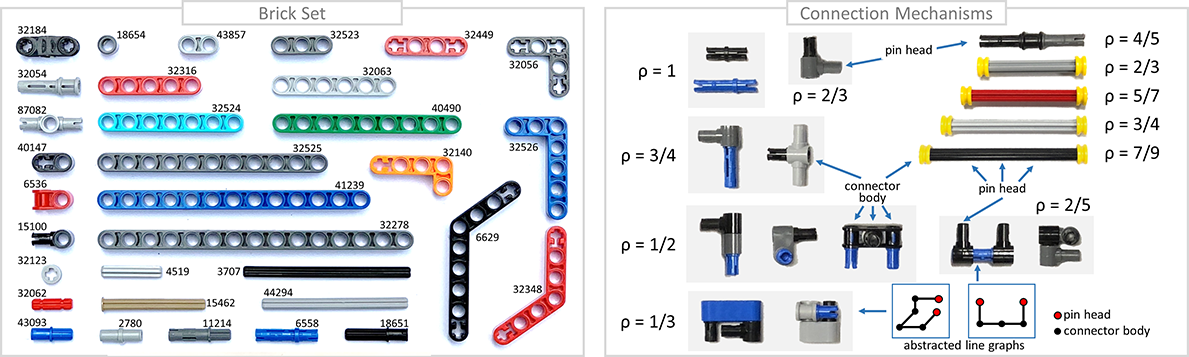}
  \vspace*{-1mm}
  \caption{\legomark Technic bricks \final{with their unique brick IDs} (left) and connection mechanisms (right) supported in our system.
The connection mechanisms are grouped by the pin head ratio ($\rho$), which is defined for quantifying the simplicity of connection mechanisms.
In addition, we show as examples the abstracted line graphs for two of the connection mechanisms.
See Supplementary material part \ed{A} for more details on the brick set and connection mechanisms.  \legoclaim}
  \label{fig:brickset_and_connections}
  \vspace*{-1mm}
\end{figure*}

To approach the problem, we first comprehensively model Technic constructions by enumerating the brick properties and connection mechanisms, conceptualizing the input as a guiding graph (see Figures~\ref{fig:overview} (a,b)), and modeling the construction requirements into an objective function.
%
\rz{Further, we formulate the brick layout problem as a {\em sketch cover\/} and solve it
%
by first estimating the local beam orientations.
Then, we further iteratively refine a random initial layout into a coherent model to cover the input sketch, while guided by the objective function; see Figures~\ref{fig:overview} (c,d).}
%

Overall, our work makes the following contributions:
\begin{itemize}
\item
\new{the first method that automatically constructs \legomark Technic models that are coherently-connected, assemblable, and functional, based on the user-provided model specifications;}
%
%
\item
a computational model for Technic constructions, considering various brick properties and types, connection mechanisms, coherency, and construction requirements, \rz{including support for joint rotations, 
a natural and fundamental mechanical functionality provided by \legomark Technic designs;} and
%
%
%
\item
finally, a working system, which enables user input designs, provides structural integrity analysis, and produces assembly instructions and visualizations of the assembly process to facilitate physical constructions (see Figures~\ref{fig:overview} (e,f,g)).
\end{itemize}


We demonstrate our method by generating Technic models of varying forms and complexities, and building physical assemblies for most of them.
We evaluate our method in various aspects, including a comparison with four baselines \rz{and with designs created by a \legomark expert.}
%
\phil{Our method took only 36 sec. to compute the {\sc airplane} model shown in Figure~\ref{fig:teaser}, while\/} more complex models can be realized in just minutes.
\rz{In contrast, it took the human expert close to 1.5 hours to design a comparable model for {\sc airplane} using existing software}.
%
%
%
Finally, we show how our method \phil{consumes motion annotations\/} and produces dynamic Technic models that exhibit hinge-style rotations and embed gear systems; see Section~\ref{sec:results}.

\section{Related works}
\label{sec:related}

\paragraph{\legomark construction.}
The first work aimed at automatic construction of \legomark models using regular bricks was by Gower et al.~\shortcite{Gower-1998-Lego}. The problem was formulated
as a combinatorial optimization to maximize a goodness measure for \legomark structures.
Some follow-up works include Petrovi\v{c}~\shortcite{Petrovic-2001-Solving} using evolutionary algorithms, Winkler~\shortcite{Winkler-2005-Automated} using beam search, Testuz et al.~\shortcite{Testuz-2013-Automatic} using a graph-based algorithm, Stephenson~\shortcite{stephenson-2016-multi-phase} using a multi-phase approach, and \new{Lee et al.~\shortcite{lee2018split} using a genetic algorithm}.
%
Considering not only the target shape, Luo et al.~\shortcite{luo-2015-legolization} developed a comprehensive method for buildable \legomark structures in larger scales, considering the brick colors in the assemblies and the structural stability.
Later, Yun et al.~\shortcite{yun-2017-LEGO-silhouette-fitted} improved \legomark constructions by silhouette fitting.
See~\cite{kim-2014-LEGO-survey} for a survey on \legomark layout methods.
%

Besides regular brick models, Lambrecht~\shortcite{lambrecht-2006-voxelization} computed \legomark\\ assemblies with oriented thin plates.
Wa{\ss}mann and Weicker~\shortcite{Wassmann-2012-stability} devised a two-phase approach for stability analysis by solving a max-flow network.
Kuo et al.~\shortcite{kuo-2015-pixel2brick} 
computed brick sculptures from pixel arts by considering visual quality and stability.

So far, existing computational works have focused on regular \legomark bricks.
In comparison, \legomark Technic designs involve a significantly more complex brick set and entirely different assembly mechanisms using a variety of pins and connectors; \new{they even support dynamic functionalities, such as joint rotations, that our current method is able to realize fully automatically}.
On the other hand, while there are a number of commodity software tools to aid users to design \legomark models, e.g., \legomark Digital designer~\cite{Lego-2018-Digital-Designer}, MLCad\shortcite{MLCad-2018}, and LDview~\shortcite{LDview-2018}, such tools only provide basic modeling and rendering for users to create \legomark\\ designs via simple drag-and-drop.
We are not aware of any advanced computational support for designing \legomark Technic models.

\vspace*{-3pt}
\paragraph{Assembly-based fabrication.}
%
%
%
Lau et al.~\shortcite{lau-2011-furniture} took a 3D man-made object as inputs, and generated parts and connectors for building the object.
%
%
Thomaszewski et al.~\shortcite{thomaszewski-2014-linkage-characters} designed motorized assemblies of linkage-based characters.
\new{Schulz et al.~\shortcite{schulz2014design} proposed a data-driven method to design 3D models.}
%
Cignoni et al.~\shortcite{cignoni-2014-mesh-joinery} generated ribbon-shaped interlocking planar slices for assembling complex 3D shapes.
Skouras et al.~\shortcite{skouras-2015-interlocking-elements} designed assemblies made up of interlocking quadrilateral elements.
Song et al.~\shortcite{song-2016-cofiFab} built 3D-printed objects by collectively using 3D-printed and laser-cut parts.
Recently, Yao et al.~\shortcite{yao-2017-decorative-joinery} created an interactive tool for designing decorative joints, while Geilinger et al.~\shortcite{geilinger-2018-Skaterbots} presented a design tool for robots that move using arbitrary arrangements of legs and wheels.

In architecture modeling, Deuss et al.~\shortcite{deuss-2014-self-supporting} minimized the assembly work for building masonry structures;
%
Yoshida et al.~\shortcite{yoshida-2015-manufacturing} devised a method for building architecture-scale models formed by glued chopsticks; 
%
Pietroni et al.~\shortcite{pietroni-2017-tensegrity} designed a computational framework for tensegrity structures; and
\final{Desai et al.~\shortcite{desai2018assembly} developed a computational design system for electromechanical devices.}


\rzf{Most previous works from this domain decompose target shapes into customized parts to facilitate 3D fabrication. In contrast, 
our goal is to reconstruct a 3D shape using a diverse but fixed (thus not customized) brick set with a rich variety of brick connections, to
approximate the target shape specifications geometrically, structurally, and in terms of dynamics, while ensuring coherence of the 
brick connections. Our problem is that of a {\em multi-tiling\/} rather than shape decomposition. Instead of accounting for various constraints 
related to 3D constructions,} \final{we must meet complex and multifaceted objectives, including simplicity, symmetry, rigidity, etc., to support the Technic 
constructions. We are not aware of any previous computational assembly works that consider an assembly of multiple brick 
types with diverse brick connections in 3D.}

\vspace*{-3pt}
\paragraph{Application of \legomark bricks.}
Lastly, we discuss applications that use \legomark bricks as off-the-shelf building elements.
In earlier works, Mitra and Pauly~\shortcite{mitra-2009-shadowart} used \legomark bricks to build shadow art sculptures and
Baronti et al.~\shortcite{baronti-2010-camera-calibration} considered \legomark structures composed of regular bricks as markers for camera calibration.
Song et al.~\shortcite{song-2012-recursive} used \legomark bricks with flat tiles to build polycube-shaped interlocking puzzle pieces, while Mueller et 
al.~\shortcite{Mueller-2014-faBrickation} developed the faBrickator system for rapidly prototyping functional objects by substituting parts of the 3D 
objects with \legomark brick assemblies. Most recently, Chen et al.~\shortcite{chen-2018-prefab} proposed to fabricate 3D objects with 3D-printable pyramidal parts to form the outer shells and universal blocks (or \legomark bricks) to build the internal cores.

\xh{
}


\section{Overview}
\label{sec:overview}


\paragraph{Input and output.}
We provide a GUI tool for sketching line segments to craft 3D \legomark Technic designs, where the line segments are constrained to have integer lengths to match Technic bricks and nearby line segment endpoints are automatically snapped to join.
Also, our tool shows semi-transparent guiding planes aligned with the principal axes for sketching coplanar lines, since Technic bricks mostly lay on the principal ($xy$, $yz$, and $zx$) planes; see Figure~\ref{fig:examples}(b).
%
%
\new{Optionally, the user may provide motion annotations to specify {\em hinge-style rotations} at joints and to indicate embedded dynamic parts in the sketch designs.}
The output of our tool includes a \legomark Technic design composed of Technic bricks, and assembly instructions and visualizations for the assembly process.

\vspace*{-3pt}
\paragraph{Objectives.}
Our method aims to optimize a combination of the following three objectives 
for the generated \legomark Technic designs:
\begin{itemize}
\item
{\em Faithfulness to input line sketch.} \
First, the output designs should resemble the input sketches and respect the intended symmetry and motion specifications provided by the user.
\item
{\em Structural integrity.} \
Second, the output should be coherently-connected and assemblable,
while striving for rigidity.
%
%
\item
{\em Simplicity and efficiency.} \
Third, we strive for an output design that is both simple and cost-effective for assembly.
\end{itemize}

\vspace*{-3pt}
\paragraph{Challenges.}
The automatic generation of \legomark Technic models involves four sub-problems:
(i) which bricks to use and where to put them; this is analogous to the set cover problem, since the bricks should cover the input design;
(ii) beam/brick orientation in the layout;
(iii) layering (see Figure~\ref{fig:layering});
and (iv) beam/brick connections.
%
%
These four sub-problems are closely coupled, thus complicating the algorithm design.
Also, we have to attentively consider the objectives during the construction process, while efficiently exploring the immense search space for finding the optimal solution, i.e., a layout of bricks that form the input design.

\paragraph{Our approach.}
First, we 
study existing Technic models~\cite{isogawa-2010-technic-idea,kmiec-2016-builder-guide}, and enumerate the brick properties and connection mechanisms.
Then, we abstract each input as a {\em guiding graph\/} \new{and each brick as a line graph} to facilitate computation (see Figures~\ref{fig:overview} (a) \& (b)).
Next, we formulate an objective to meet the various construction requirements, \new{develop the layout modification operator to locally update brick layouts, and design a beam connection procedure to join adjacent beams.}
%
%
\phil{Further, we formulate the layout search as a sketch cover to automate Technic constructions\/}:
\begin{itemize}
%
\vspace*{-1mm}
\item
For simple connections and structural integrity, Technic beams in local structures often have the same orientation.
Hence, we first estimate the local orientation of beams over the design for sub-problem (ii) in the first stage; see Figure~\ref{fig:overview}(c).
%
\item
Next, in the second stage, \new{we optimize mainly for sub-problems (i), (iii) and (iv) altogether by first initializing a random layout}, then formulating an iterative procedure to refine the layout and connect beams, guided by the objectives; see Figure~\ref{fig:overview}(d).
%
\end{itemize}

\section{Modeling \legomark Technic Construction}
\label{sec:modeling}





\begin{figure}[!t]
  \centering
  \includegraphics[width=8.35cm]{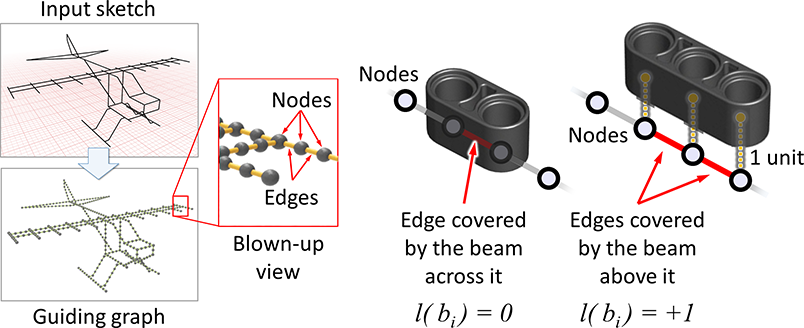}
  \vspace*{-1mm}
  \caption{An example guiding graph (left) and edge cover (right).}
  \label{fig:guiding-graph}
  \vspace*{-1mm}
\end{figure}


\subsection{Technic Bricks}
\label{ssec:model-bricks}


\paragraph{Brick set.}
%
Figure~\ref{fig:brickset_and_connections} (left) shows the bricks supported by our system.
For each brick, our system stores its 3D mesh model, physical properties such as weight, and connecting locations on the brick, i.e., {\em holes\/} on beams and connectors, and {\em pin heads\/} on pins and axles; see Figure~\ref{fig:brickset_and_connections} (right) for examples.
Also, we abstract the structure of each brick as a simple {\em line graph\/} in 3D.
For example, the green beam in Figure~\ref{fig:brickset_and_connections} (left) is abstracted as a graph of nine nodes (holes) and eight one-unit-long edges between adjacent nodes.
%
%


\vspace*{-3pt}
\paragraph{Brick connections.}
Next, we enumerate the beam connection mechanisms \final{by studying the connection mechanisms in existing Technic models}, where pin heads are locations that connect to beam holes, and connector bodies are non-pin-head locations in the connection mechanisms; see the labels in Figure~\ref{fig:brickset_and_connections} (right).
Also, we abstract each mechanism as a line graph in 3D; see Figure~\ref{fig:brickset_and_connections} (right) for two examples.
We consider two types of pin heads, i.e., {\em axle\/} and {\em regular\/}, for 
\setlength{\columnsep}{2.5mm} 
\begin{wrapfigure}{r}{0.34\columnwidth}
	\vspace{-11pt}
	\includegraphics[width=0.33\columnwidth]{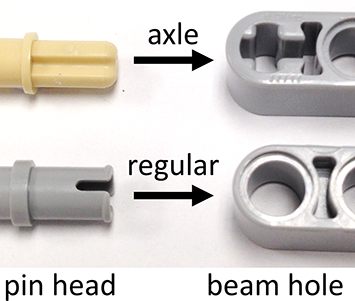}
	\hspace{-30pt}
	\vspace{-10pt}
\end{wrapfigure}
connecting respective types of beam holes; see the right inset figure.
The two types tradeoff between connection rigidity and flexibility: the axle pin heads enforce rigid non-rotatable connections that must be perpendicular/parallel, while the regular pin heads allow rotatable connections.


\vspace*{-3pt}
\paragraph{Pin-head ratio, $\rho$.}
In \legomark Technic, simpler connection mechanisms dominated by pin heads (not connector bodies) are preferred, since they help enhance the structural integrity.
To quantify the simplicity of connection mechanisms, we define $\rho$ as the pin head count over the total node count in a mechanism's line graph, where simpler mechanisms have larger $\rho$ values; see Figure~\ref{fig:brickset_and_connections} (right).


%

\begin{figure*}[!t]
	\centering
	\includegraphics[width=17.7cm]{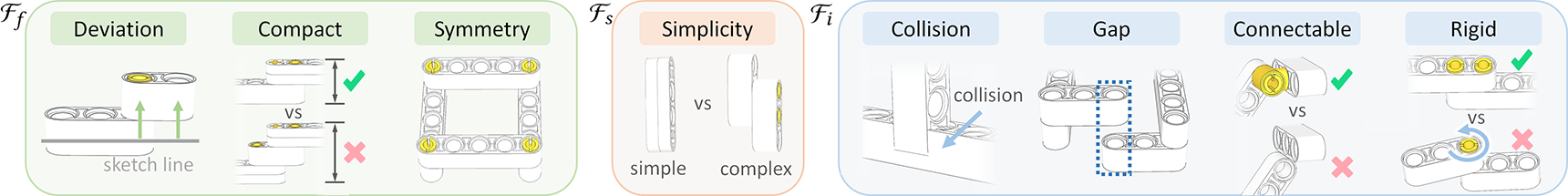}
	\vspace*{-3mm}
	\caption{\new{Component terms in the objective function, from left to right:
faithfulness to input sketch ($\mathcal{F}_f$), model simplicity ($\mathcal{F}_s$), and structural integrity ($\mathcal{F}_i$).}}
	\label{fig:obj_func}
	\vspace*{-1mm}
\end{figure*}


\vspace*{-3pt}
\paragraph{Layer number, $l(b_i)$.}
Technic beams of the same orientation are connected in layers, above or below one another; see Figure~\ref{fig:layering}.
Given a beam, say $b_i$, we define its {\em layer number\/} in a \legomark Technic construction as $l(b_i)$, where $l(b_i)$ is zero, if $b_i$ exactly goes through the associated sketch line in the input, and $l(b_i)$ is positive/negative, if $b_i$ is above/below the associated sketch line; see Figure~\ref{fig:guiding-graph} (right).



\subsection{The Sketch Cover problem}
\label{ssec:model-sketch}



\paragraph{Guiding graph.}
For efficient computation, we abstract the input sketch as a {\em guiding graph\/} (denoted as $\mathcal{G}$), where nodes are distributed along the sketch line segments with adjacent nodes being one unit apart, like holes on Technic beams.
Moreover, edges in $\mathcal{G}$ are all one unit long for connecting the adjacent nodes; see Figure~\ref{fig:guiding-graph} (left).



\vspace*{-3pt}
\paragraph{Sketch cover.}
An edge in $\mathcal{G}$ is said to be {\em covered\/} by a brick (e.g., beam) if part of the brick is parallel to it, either crossing it or locating at a small distance from the edge; see Figure~\ref{fig:guiding-graph} (right).
If all edges in $\mathcal{G}$ are covered by some bricks in a generated Technic model, the model is said to {\em fully cover\/} the input sketch.
\new{The {\em sketch cover problem\/} is to find a brick layout that fully covers the given input design.\/}



\vspace*{-3pt}
\paragraph{Symmetric groups.}
Our system considers reflection and translational symmetry.
We group line segments (i.e., subgraphs in guiding graph) and mark symmetric groups in the input sketch.
In our current implementation, this is done manually, but in the future, we plan to incorporate symmetry detection methods for the task.

\subsection{Objective function}
\label{ssec:obj}

\final{

We formulate and minimize the following objective function to guide the finding of a beam layout (say $\mathcal{B} = \{ b_i \}$) that covers the input sketch, while conforming to the various construction requirements:
%
%
\begin{equation}
\nonumber
	\label{eq:obj}
	\mathcal{F}
	\ = \
	\mathcal{F}_f
	+
	\mathcal{F}_s
	+
	\mathcal{F}_i \ ,
\end{equation}
where $\mathcal{F}_f$, $\mathcal{F}_s$, and $\mathcal{F}_i$ are component terms in the objective (see the corresponding illustrations shown in Figure~\ref{fig:obj_func}):
%


\vspace*{-3pt}
\paragraph{(i) Faithfulness to input sketch, $\mathcal{F}_f$.}
We define
\begin{equation}
\nonumber
\mathcal{F}_f
\ = \
w_\text{dev} \mathcal{F}_\text{dev}
+
w_\text{cpt} \mathcal{F}_\text{cpt}
+
w_\text{sym} \mathcal{F}_\text{sym} \ ,
\end{equation}
where $w_\text{dev}$, $w_\text{cpt}$, and $w_\text{sym}$ are weights, and we have:
\begin{itemize}
\item
$\mathcal{F}_\text{dev}$ minimizes the distance deviation (i.e., layer number $l(b_i)$) of the beams in the layout from the input sketch:
\begin{equation}
\nonumber
\mathcal{F}_\text{dev}
\ = \
\sqrt{\frac{1}{\sum_{b_i \in \mathcal{B}} L(b_i)} \ \sum_{b_i \in \mathcal{B}} L(b_i) \ l(b_i)^2} \ ,
\vspace*{-1mm}
\end{equation}
where $L(b_i)$ denotes the length (number of holes) of beam $b_i$.
\item
$\mathcal{F}_\text{cpt}$ compacts the layering by minimizing the range of $l(b_i)$ in each coplanar component (see Section~\ref{ssec:beam_orient}) in the layout:
\begin{equation}
\nonumber
\label{eq:cpt}
\mathcal{F}_\text{cpt}
\ = \
\max_{\mathcal{C}_j \in \mathcal{C}} \ [ \ \max_{b_i \in \mathcal{C}_j} l(b_i) \ - \ \min_{b_i \in \mathcal{C}_j} l(b_i) \ ]^2 \ ,
\vspace*{-1mm}
\end{equation}

where $\mathcal{C} = \{ \mathcal{C}_j \}$ is a set of coplanar components extracted from the input design at the end of the stage one of our framework, and each $\mathcal{C}_j$ is a subset of beams in $\mathcal{B}$; see Section ~\ref{ssec:beam_orient} for how we extract $\mathcal{C}$, the set of coplanar components.
\item
$\mathcal{F}_\text{sym}$ minimizes the deviation from symmetry for each pair of symmetric beam groups:
\begin{equation}
\nonumber
\label{eq:sym}
\mathcal{F}_\text{sym}
\ = \
\frac{1}{|\mathcal{S}|}
\sum_{(\mathcal{B}_i,\mathcal{B}_j,T_{ji}) \in \mathcal{S}} \ \frac{1}{\sum_{b \in \mathcal{B}_i} L(b)} \ d_H(\mathcal{B}_i,T_{ji}(\mathcal{B}_j))^2 \ ,
\end{equation}
where
$\mathcal{B}_i \subset \mathcal{B}$ and $\mathcal{B}_j \subset \mathcal{B}$ are symmetric groups;
$T_{ji}$ is the 3D symmetric transformation to bring $\mathcal{B}_j$ to $\mathcal{B}_i$;
$\mathcal{S}$ is a set of symmetric groups; and
$d_H$ denotes the Hausdorff distance.
In the case of self (reflection) symmetry, we can have $\mathcal{B}_i = \mathcal{B}_j$.
\end{itemize}


\vspace*{-3pt}
\paragraph{(ii) Model simplicity, $\mathcal{F}_s$.}
We encourage {\em layout simplicity\/} by minimizing the total beams length and maximizing the total pin head ratios of the connection mechanisms in the layout;
\begin{equation}
\nonumber
\label{eq:simplicity}
\mathcal{F}_s
\ = \
w_\text{tbl} \frac{\sum_{b_i \in \mathcal{B}} L(b_i)}{|\mathcal{V}|}
+
w_\text{phr} (1-\bar{\rho}) \ ,
\end{equation}
where
$\mathcal{V}$ is the set of nodes in guiding graph $\mathcal{G}$;
$\bar{\rho}$ is the {\em average pin-head ratio\/} (phr) over all the connections in the \legomark Technic model; and
$w_\text{tbl}$ and $w_\text{phr}$ are weights.
%


\begin{figure*}[!t]
  \centering
  \includegraphics[width=17cm]{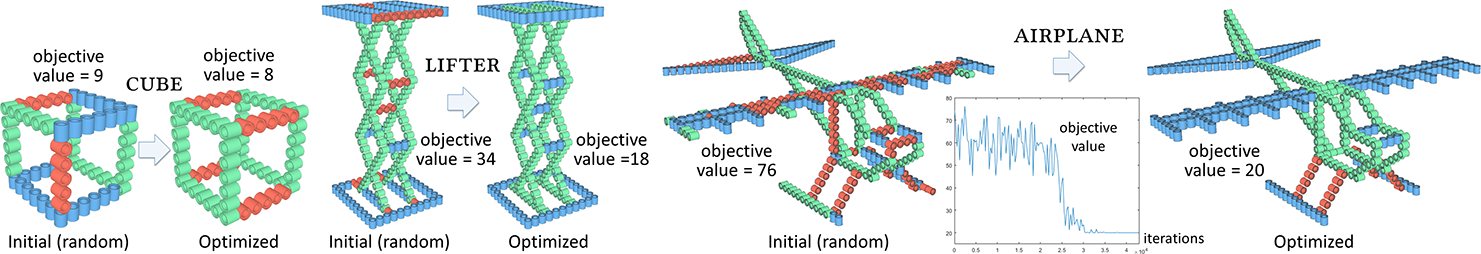}
  \vspace*{-2mm}
  \caption{Optimizing the ``beam hole'' orientations, such that most adjacent beam holes have the same orientation for simpler beam connections.}
  \label{fig:optimize_SA}
  \vspace*{-1mm}
\end{figure*}


\vspace*{-3pt}
\paragraph{(iii) Structural integrity, $\mathcal{F}_i$.}
We define:
\begin{equation}
\nonumber
\mathcal{F}_i
\ = \
w_\text{col} \mathcal{F}_\text{col}
+
w_\text{gap} \mathcal{F}_\text{gap}
+
w_\text{coh} \mathcal{F}_\text{coh}
+
w_\text{rgd} \mathcal{F}_\text{rgd} \ ,
\end{equation}
where $w_\text{col}$, $w_\text{gap}$, $w_\text{coh}$, and $w_\text{rgd}$ are weights, and we have:
\begin{itemize}

\item
$\mathcal{F}_\text{col}$ measures the total number of collisions ($N_{col}$) between beams in layout $\mathcal{B}$ and normalizes it by the total beam length:
\begin{equation}
\nonumber
\mathcal{F}_\text{col}
\ = \
\frac{1}{\sum_{b_i \in \mathcal{B}} L(b_i)} \ N_\text{col}(\mathcal{B}) \ .
\end{equation}

\item 
$\mathcal{F}_\text{gap}$ measures the total number of gaps between beams.
We denote $\mathcal{B}_v$ as the set of beams that associate with vertex $v \in \mathcal{V}$.
Then, we detect the gap at $v$ by
\begin{equation}
\nonumber
\text{gap}(v)
\ = \
[ \ \max_{b_i \in \mathcal{B}_v}l(b_i) - \min_{b_i \in \mathcal{B}_v}l(b_i) \ ] + 1 - |\mathcal{B}_v| \ ,
\vspace*{-1mm}
\end{equation}
%
and define $\mathcal{F}_\text{gap} \ = \ \frac{1}{|\mathcal{V}|} \ \sum_{v \in \mathcal{V}} \text{gap}(v) \ $.
Although we may fill a gap using a single hole brick (ID 18654 in Figure~\ref{fig:brickset_and_connections}), but we still need to reduce gaps for structural integrity.

\item 
$\mathcal{F}_\text{coh}$ measures the connection coherence (connect-ability) between beams in $\mathcal{B}$.
To start, we first connect adjacent beams in $\mathcal{B}$ by trying various connection mechanisms (see Section ~\ref{ssec:key_components}), and count the number of failure connections ($N_\text{cfail}$).
%
%
We define $\mathcal{F}_\text{coh} = \frac{N_\text{cfail}}{|\mathcal{E}|}$, where $\mathcal{E}$ is the edge set in guiding graph.
Note that if we fail to connect two adjacent beams, the layering and/or orientation of the beam(s) will be modified in the iterative refinement process; see Section ~\ref{ssec:layout_refine} for detail.
%

\item
$\mathcal{F}_\text{rgd}$ measures the number of rigid beam subsets in the layout.
In general, a local beam connection can be rigid or non-rigid; see Figure~\ref{fig:obj_func} (right).
Here, we perform a depth first traversal over the guiding graph to examine the connections between adjacent beams, and stop the traversal at non-rigid connections.
As a result, we can decompose $\mathcal{B}$ into ``rigid'' subsets, such that all the beams inside are rigidly connected transitively.
In this way, we can evaluate
\begin{equation}
\nonumber
\label{eq:rgd}
\mathcal{F}_\text{rgd}(\mathcal{B})
\ = \
\frac{\text{number of rigid subsets in} \ \mathcal{B}}{\sum_{v \in \mathcal{V}} \ (\text{deg}(v) - 1)^2} \ ,
\end{equation}
where $\text{deg}(v)$ is the valence of vertex $v$ in $\mathcal{G}$.
Note that we also tried to normalize the term by using the number of beams in $\mathcal{B}$ instead, but we find the above formulation to be more scalable to various input designs, since it measures not only the scale but also the topological complexity.


\end{itemize}

To balance the components in the objective, we empirically set the associated weights as follows: $w_\text{cpt}$, $w_\text{sym}$, $w_\text{phr}$, $w_\text{col}$, $w_\text{gap}$, and $w_\text{coh}$ are set as $1$, $30$, $10$, $30$, $100$, and $50$, respectively. 
On the other hand, $w_\text{dev}$, $w_\text{tbl}$, and $w_\text{rgd}$ are set based on user preference. In practice, we set $\{ w_\text{dev}, w_\text{tbl}, w_\text{rgd} \}$ as $\{100,0,0\}$ to aim for high faithfulness to the input sketch, as $\{0,100,0\}$ to aim for model simplicity, and as $\{100,0,100\}$ or $\{0,100,100\}$ to additionally aim for rigid connections. Figure~\ref{fig:preference} and Table~\ref{fig:abla_study} show experiments on their effectiveness.

}

\section{\legomark Technic Construction Search}
\label{sec:search}

\new{In our initial attempts, we tried a greedy approach that progressively arranges locally-optimum beams to cover the sketch.
Further, we tried several other approaches (see method comparisons in Section~\ref{sec:results}) to improve the search, but the results produced from these approaches are poorly optimized, especially for nontrivial inputs.
To address the immense search space (combinations of beam placements, orientations, connections, and layering) in Technic constructions, the search has to be efficient and allow updates that iteratively propagate over the layout.
Hence, we design a two-stage approach (see Figures~\ref{fig:overview}(c) \& (d)) that first estimates the beam orientation then iteratively refines the layout to optimize the objectives.\/}

%


\subsection{Stage one: Estimate Beam Orientation}
\label{ssec:beam_orient}

In \legomark Technic models, adjacent beams of same orientation can be steadily connected by pins.
However, to build 3D models, we generally need to arrange beams in different orientations for building different parts of the models.
Since the beam orientation strongly affects the overall structure, connections \phil{and joint rotations\/}, we first estimate the local beam orientation over the guiding graph.

\begin{wrapfigure}{r}{0.37\columnwidth}
	\vspace{-10pt}
	\hspace{-5pt}
	\includegraphics[width=0.36\columnwidth]{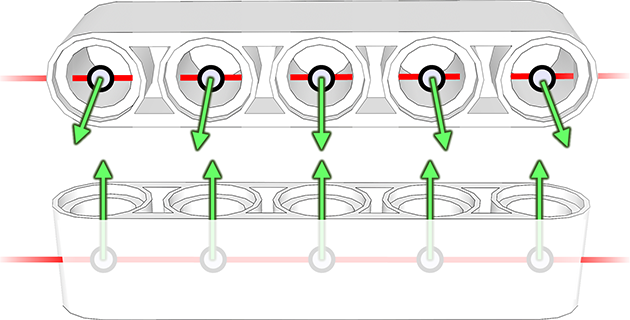}
	\hspace{-30pt}
	\vspace{-10pt}
\end{wrapfigure}
Mathematically, we represent a beam's orientation as a 3D vector that passes through the medial axis of the beam's holes; see the inset figure.
In general, we can reorient a beam, as long as its orientation vector is perpendicular to either its corresponding line segment in the sketch, or the edges that it covers in the guiding graph.
%
%
%
%
Here, we model the problem of finding the beam orientation as an assignment problem.
Since most nodes in guiding graph $\mathcal{G}$ will eventually be covered by beams in the generated \legomark Technic model (the rest will be covered by the connection mechanisms), we create a {\em hole orientation variable\/} for each node in $\mathcal{G}$.
Since \legomark Technic bricks mostly lay on the principal (xy, yz and zx) planes, we should assign to each variable a principal direction (X, Y or Z), unless purposely specified in the user interface.


\vspace*{-3pt}
\paragraph{Goal.}
To {\em minimize the number of adjacent node pairs with different hole orientations\/}, such that we can encourage the use of pin and axle connections for model simplicity and structural integrity.


\vspace*{-3pt}
\paragraph{Constraints}
(i) For non-junction nodes along line segments in the input sketch, their hole orientations should be perpendicular to the associated line segment~\footnote{\setlength{\columnsep}{3.5mm}
%
\begin{wrapfigure}{r}{0.27\columnwidth}
	\vspace{-20pt}
	\hspace{-5pt}
	\includegraphics[width=0.26\columnwidth]{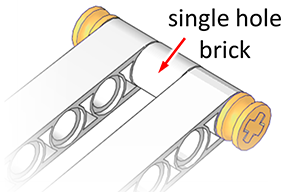}
	\hspace{-30pt}
	\vspace{-15pt}
\end{wrapfigure}
Except for short line segments with just a single interior hole (see the right inset figure), since we observe that in such a situation, existing \legomark models may fill/cover the node using a single hole brick (near Figure~\ref{fig:brickset_and_connections} (top-left)) whose orientation aligns with the associated line segment.};
%
(ii) orientation of adjacent holes should be the same or perpendicular to each other, so that we may use a single beam to cover the two holes or connect them using a connection mechanism shown in Figure~\ref{fig:brickset_and_connections} (right); 
\phil{(iii) at each joint annotated to allow rotation, the orientation of the associated hole must align with the joint's rotational axis; and\/}
(iv) hole orientation variables at symmetry locations are constrained to be the same (for translational symmetry) or mirrored (for reflection symmetry).

\vspace*{-3pt}
\paragraph{Method.}
We solve this combinatorial optimization problem using a simulated annealing model\final{~\cite{kirkpatrick1983optimization}}.
Initially, we randomize all hole orientation variables (see Figure~\ref{fig:optimize_SA} for examples) but following the listed constraints.
Then, we iteratively choose a random line segment, change the orientation of all the non-junction nodes in the segment, and update the orientation at each associated junction node, if the objective is minimized.
Figure~\ref{fig:optimize_SA} shows the initial state, optimized result, and objective values for the three examples.
In our implementation, we set the initial temperature as $2\times10^3$, stopping temperature as $0.01$, and cooling rate as $1-\frac{10}{|\mathcal{V}|}$, where $\mathcal{V}$ is the node set in the guiding graph.
The optimization completes in only a few seconds for most models; see Section~\ref{sec:results}.
%


\begin{figure}[!t]
\centering
\includegraphics[width=0.9\linewidth]{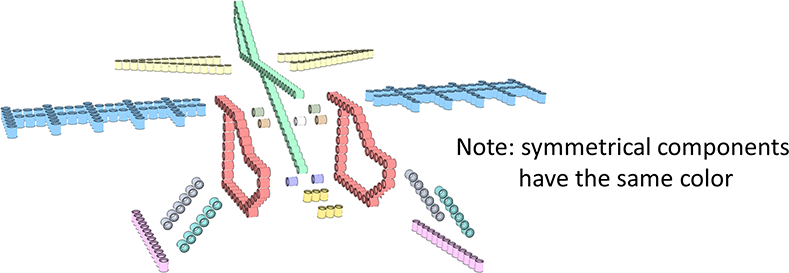}
\vspace*{-2mm}
\caption{The optimized {\sc airplane} in Figure~\ref{fig:optimize_SA} has 22 components. \legoclaim}
\label{fig:plane_components}
\vspace*{-2mm}
\end{figure}

\begin{figure}[!t]
\centering
\includegraphics[width=0.95\linewidth]{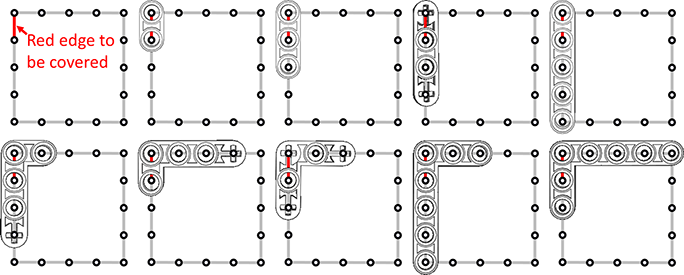}
\vspace*{-2mm}
\caption{Nine possible beam placements to cover the red edge \final{(see top left)\/}.}
\label{fig:beam_placement}
\vspace*{-2mm}
\end{figure}



\vspace*{-3pt}
\paragraph{Find ``components'' and ``beam placements''}
Next, we decompose nodes in guiding graph into \final{{\em connected\/} and {\em coplanar\/}} components, where the nodes in each component \final{have the {\em same orientation\/}}.
%
%
For example, the optimized {\sc cube} and {\sc lifter} shown in Figure~\ref{fig:optimize_SA} have six and nine components, respectively, while the optimized {\sc airplane} has 22 components; see Figure~\ref{fig:plane_components}.
Also, we store symmetry information within and between components to facilitate later computation.

Furthermore, on each component, \final{we find all feasible beam placements on the component}, but ignoring the beam placements that pass through the annotated rotating joints.
Hence, for each edge in a component, we keep {\em a list of feasible beam placements\/} that can cover the edge (see Figure~\ref{fig:beam_placement} for an example), and also, a list of beam placements that stop at or pass through each node.
\phil{Using this data structure, we can efficiently arrange beams to cover any edge in the guiding graph, and accelerate the layout generation in the second stage (layout modification operator) of our method.\/}


\subsection{Key components in Stage Two}
\label{ssec:key_components}

\new{
\paragraph{Layout modification operator.}
\final{There are two key components in the second stage of our method.
The first one is an operator to modify a given beam layout.
We design this operator\/} with the following considerations:
(i) the operator must be {\em efficient\/}, due to its heavy usage in the search process;
(ii) even if the operator is local, successively applying it should produce {\em diverse beam layouts\/}; and
(iii) it should {\em avoid obviously bad beam placements\/}. 

Figure~\ref{fig:operator} shows the operator procedure.
After randomly picking a covered edge in the guiding graph, we remove all the beams that stop at or pass through the edge, and locate all possible beam placements that can cover the resulting ``uncovered'' edges in the guiding graph.
To promote layout diversity, \final{we next calculate a selection probability for each possible beam placement, where each beam candidate is selected based on the number of uncovered edges that it can cover;}
see examples in Figure~\ref{fig:operator}.
Next, based on the probabilities, we randomly select a candidate beam placement to add to the layout, at a layer that produces more compact layering.
Hence, we may try both long and short beams in the search, while avoiding meaningless beams that \final{cannot\/} cover any edge and encouraging model simplicity and faithfulness to the \final{input\/} sketch.
We repeat this select-and-add process (usually a few times) until we cover all the uncovered edges.
%
%
This procedure was carefully designed after experimenting with several alternatives; see Supplementary material part \final{C}.

\begin{figure}[!t]
  \centering
  \includegraphics[width=8.35cm]{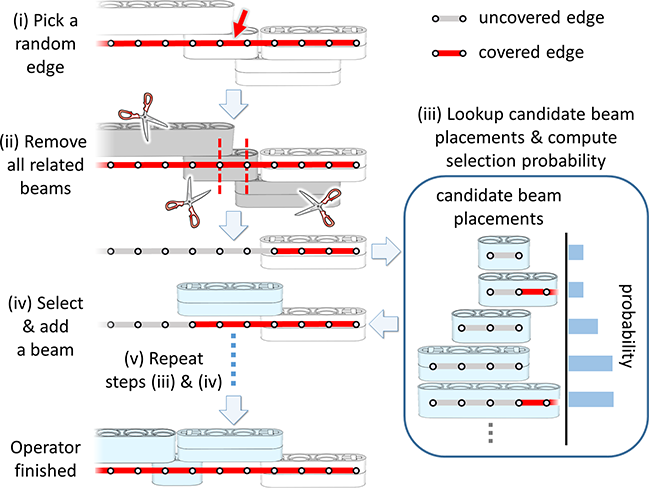}
  \vspace*{-2mm}
  \caption{\new{The layout modification operator efficiently modifies a layout by locally removing beams around a random edge and adding new beams.}}
  \label{fig:operator}
  \vspace*{-2mm}
\end{figure}

\begin{figure}[!t]
\centering
\includegraphics[width=7.2cm]{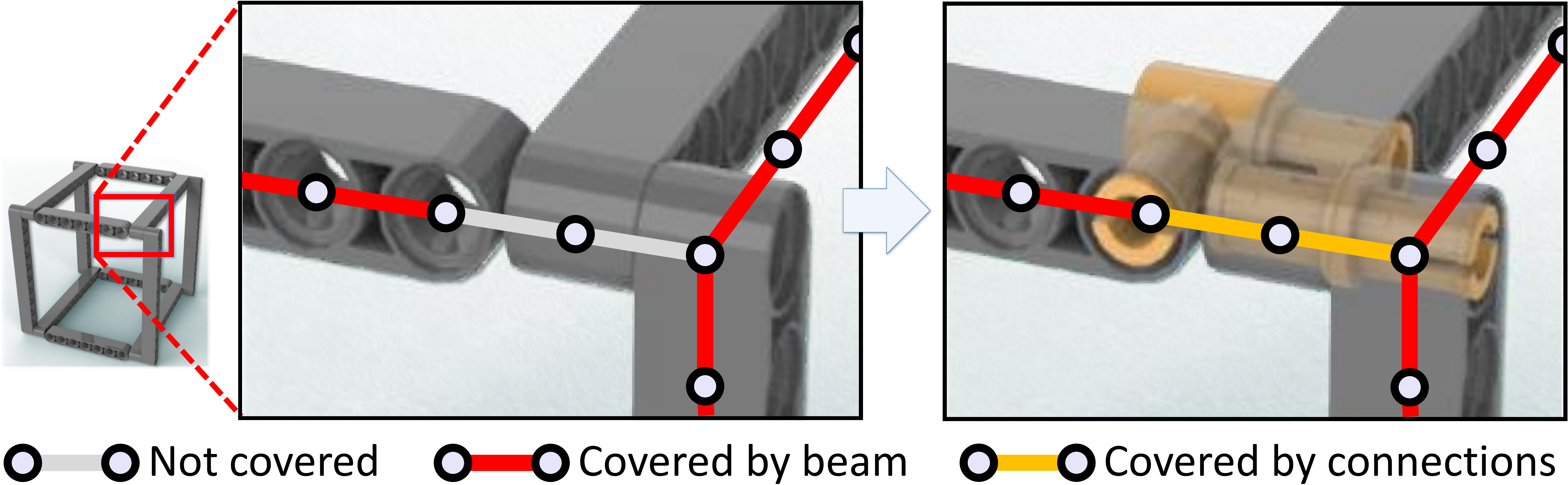}
\vspace*{-2mm}
\caption{After we arrange the beams (left), we need to further arrange connection mechanisms (in orange) between adjacent beams (right).}
\label{fig:example_conn_cover}
\vspace*{-2mm}
\end{figure}

\begin{figure}[!t]
	\centering
	\includegraphics[width=8.35cm]{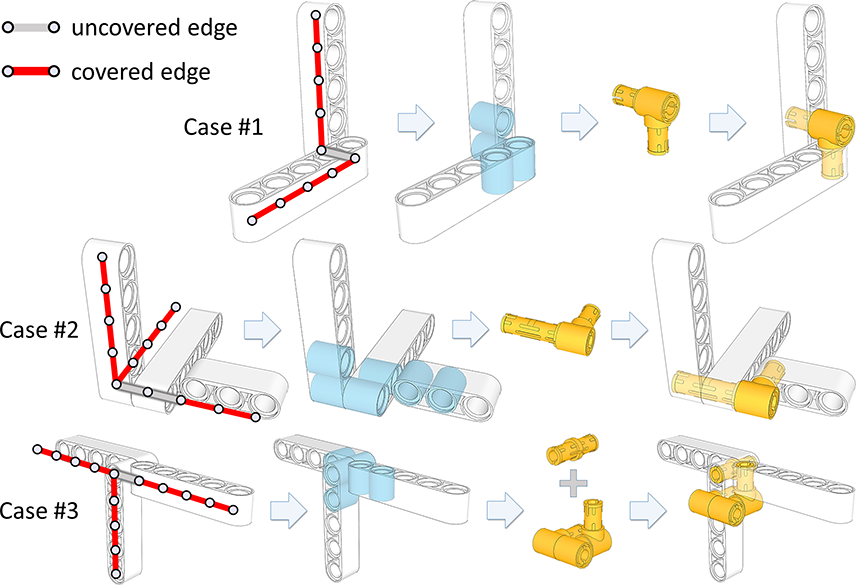}
	\vspace*{-2mm}
	\caption{Procedure: find connection mechanisms to join adjacent beams.}
	\label{fig:running_connection}
	\vspace*{-2mm}
\end{figure}


\vspace*{-3pt}
\paragraph{Beam connection procedure.}
Another key component is the procedure to connect adjacent beams, such that we can join the beams and form a \final{connected assembly} in the end.
%
Procedure-wise, given a layout of beams, we first identify edges in the guiding graph that are not covered by any beam, and locally group the adjacent uncovered edges; see the examples shown in Figure~\ref{fig:running_connection} (leftmost column).
Then, we identify the beam holes around each group based on a distance threshold of $\sqrt{3}$ units; see the highlighted holes shown in Figure~\ref{fig:running_connection} (middle column).
To connect the beams in each group, we first find {\em all feasible connection mechanisms} that can join holes of different beams, then find {\em subsets of them\/} (see Figure~\ref{fig:running_connection} (rightmost column)) that satisfy the following considerations:
(i) the subset of mechanisms should together connect all the different beams around the uncovered edges;
(ii) the chosen mechanisms should not collide with the existing beams and also one another;
(iii) we aim for \final{maximal} pin-head ratio for model simplicity;
(iv) connecting bricks that extend outward should not go below the ground plane or collide with any embedded dynamic element; and
(v) the chosen mechanisms should not interfere any annotated joint rotations.


In general, beam connections are not always one-to-one, i.e., see the case in Figure~\ref{fig:running_connection} (top).
First, some mechanisms can join three or more beams together, e.g., the L-shaped mechanism in Figure~\ref{fig:running_connection} (middle) and the T-shaped one with $\rho$ = $3/4$ in Figure~\ref{fig:brickset_and_connections} (right).
Second, we sometimes need more than one connection mechanisms to join the beams around a group of uncovered edges, e.g., for the cases shown in Figure~\ref{fig:running_connection} (bottom) and Figure~\ref{fig:example_conn_cover}, we need two mechanisms to connect the beams.
Furthermore, in case feasible connection mechanisms cannot be found, the connectable term in the objective (see Figure~\ref{fig:obj_func}) will reflect the result, so that the search framework will be guided to modify the layout accordingly.
}




\subsection{Stage two: Iterative Layout Refinement}
\label{ssec:layout_refine}

Overall, our solution search in Stage two starts by initializing a random layout of beams, then iteratively modifies it to improve the objective function; see Figure~\ref{fig:overview}(d).
Particularly, the layout starts without beam connections, so we have to find appropriate connection mechanisms to join the beams during the search.
We have tried various optimization frameworks to optimize the objective function (see Figure~\ref{fig:comparison}), and in the end, adopted a simulated annealing model proposed by Cagan et al.~\shortcite{cagan1998simulated} to regulate the solution search.

\vspace*{-3pt}
\paragraph{Layout initialization}
We create the initial layout by repeating the first two steps in the layout modification operator, i.e., randomly pick an uncovered edge in the guiding graph and select a feasible beam placement to add into the initial layout.
However, we deliberately select beam placements with equal probability to generate a more random initial layout (see Figure~\ref{fig:overview}(d) for an example), since a more random layout helps the annealing process avoid local minima.

\vspace*{-3pt}
\paragraph{Overall procedure}
Algorithm~\ref{alg:SA} outlines the search procedure.
There are four input parameters, $T_\text{max}$, $T_\text{mid}$, $T_\text{min}$ and $r$, which denote the starting temperature, middle cutoff temperature, ending temperature, and cooling rate, respectively.
We empirically set $T_\text{max}$ as $2$$\times$$10^3$, $T_\text{mid}$ as $10$, $T_\text{min}$ as $10^{-4}$, and $r$ as $0.999$ for simple models and as $0.99997$ for large complex models to trade off model quality and running time.
In the early annealing process, the layout is highly random and not stable, so we guide the layout refinement by minimizing a simplified version of the objective function (denoted as $\mathcal{F}^0$) without evaluating the collision and connectable terms for computational efficiency, then switching to the full version objective ($\mathcal{F}$) when the layout becomes stable.
Also, since the layout modification operator is local, we actually update the objective function value based on the local changes in the layout.
This can boost the computational efficiency for evaluating the objective function.


\begin{algorithm}[!t] 
\SetAlgoNoLine
\KwData{$T_\text{max}$, $T_\text{min}$, $T_\text{mid}$, $r$, and Guiding graph $\mathcal{G}$} 

$\mathcal{B}_\text{current} = \ \text{initialize}\_\text{layout}(\mathcal{G})$ \hspace*{3mm} \tcp{\footnotesize initialize the beam layout} 
$T \hspace*{6.95mm} = \ T_\text{max}$                                          \hspace*{20.25mm} \tcp{\footnotesize initialize temperature T}
$\mathcal{B}_\text{best} \hspace*{2.9mm} = \ \mathcal{B}_\text{current}$       \hspace*{16.75mm} \tcp{\footnotesize initialize the best layout}
$\mathcal{F}_\text{obj} \hspace*{4.05mm} = \ \mathcal{F}^o$                    \hspace*{21.75mm} \tcp{\footnotesize use approx. obj. func.}
\While{ \ $T \ > \ T_\text{min}$ \ }{
    \If{ \ $T \ < \ T_\text{mid}$ \ }{
        $\mathcal{F}_\text{obj} \ = \ \mathcal{F}$                             \hspace*{17.25mm} \tcp{\footnotesize switch to full obj. func.}
    }
	$\mathcal{B}_\text{new} = \ \text{modify}(\mathcal{B}_\text{current})$ \hspace*{4.75mm} \tcp{\footnotesize layout modification op.}
	$\Delta \ \hspace*{3.6mm} = \ \mathcal{F}_\text{obj}( \mathcal{B}_\text{new} ) \ - \ \mathcal{F}_\text{obj}( \mathcal{B}_\text{current} )$\\
	\If{\ $\exp(-\Delta / T ) \ > \ \text{rand}(0,1) $ \ }{
		$\mathcal{B}_\text{current} \ = \  \mathcal{B}_\text{new}$	\hspace*{9.25mm} \tcp{\footnotesize accept the change}
		\If{ \ $\mathcal{F}_\text{obj}(\mathcal{B}_\text{current}) \ > \ \mathcal{F}_\text{obj}(\mathcal{B}_\text{best}) $ \ }{
				$\mathcal{B}_\text{best} \ = \ \mathcal{B}_\text{current}$	\hspace*{4.5mm} \tcp{\footnotesize update the best layout}
			}
	}
	$T \ = \ T*r$  \hspace*{21mm}   \tcp{\footnotesize update temperature $T$}
}
\textbf{return} \hspace*{1mm} $\mathcal{B}_\text{best}$
\caption{Overall procedure for iterative layout refinement} 
\label{alg:SA} 
\end{algorithm}




%


\vspace*{-3pt}
\paragraph{Discussion}
The beam orientations estimated in Stage one may not always be perfect; we further allow the layout modification operator to try different valid beam orientations (see the constraints in Section~\ref{ssec:beam_orient}), when the layout becomes stable.
See Figure~\ref{fig:optimize_SA} (right) for the estimated orientation of the pontoon beams on the bottom of {\sc airplane}; re-orientating them allows simpler connections (see Figure~\ref{fig:teaser}) that minimize the objective.
Besides the exponential annealing schedule we adopted in the search, we have tried other schedules: linear, logarithmic, optimum~\cite{nourani1998comparison}, and thermodynamic~\cite{de2003placement}.
However, we found no obvious improvements in results and running time.
This is likely because the search space is discrete rather than continuous, where the feasible solutions are far from one another.
Also, we have tried to extend Algorithm~\ref{alg:SA} to be a population-based search~\cite{van2007population} by finding $N$ instead of one solutions in each iteration and keeping the best $K$ for generating candidates in the next iteration (we set $N$ = 20 and $K$ = 4).
However, the solution quality improves only slightly but the running time increases substantially, so we kept $N=K=1$ when producing our results.

\begin{figure}[!t]
	\centering
	\includegraphics[width=8cm]{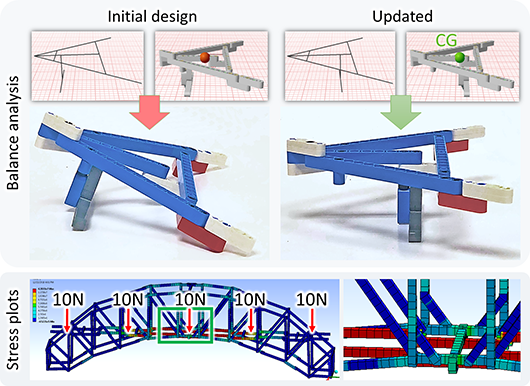}
	\vspace*{-2mm}
	\caption{\new{Example balance analysis (top) and stress plots (bottom).}  \legoclaim}
	\label{fig:balance_stress}
	\vspace*{-1mm}
\end{figure}

\begin{figure*}[t]
	\centering
	\includegraphics[width=17.8cm]{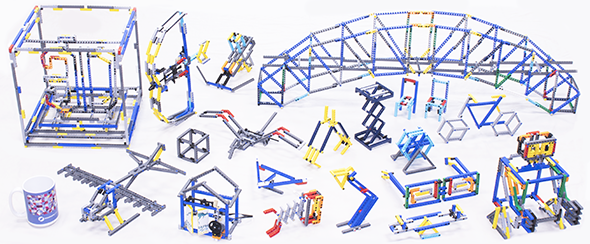}
	\vspace*{-1mm}
	\caption{A photograph showing the physical assemblies of most \legomark Technic models generated by our method.
\final{From left to right, we have\/}
{\sc print\_box},
{\sc airplane},
{\sc long\_bow},
{\sc cube},
\new{{\sc house}},
{\sc crossbow},
\new{{\sc bird}},
{\sc bridge},
{\sc flying\_kite}, 
{\sc claw},
{\sc picker},
\new{{\sc table\_lamp}}, 
{\sc lifter},
\new{{\sc chair\_front}}, 
\new{{\sc chair\_side}}, 
\new{{\sc little\_ferris}}, 
\new{{\sc glasses}}, 
\new{{\sc seasaw}}, 
\new{{\sc bicycle}}, 
and 
\new{{\sc robot}}; see Table~\ref{fig:statistics} for the statistics \final{(number of input sketch lines, bricks, etc.)\/} about these generated models. \legoclaim}
%
	\label{fig:phys_gallery}
	\vspace*{-1mm}
\end{figure*}


\subsection{Model Analysis and Assembly}
\label{ssec:postprocessing}

Our tool provides further analysis on the generated Technic model:
(i) self-balancing --- check if the model's center of gravity is well-supported\final{~\cite{Prevost-2013-make-it-stand,schneider-2002-geometric-tools,mcghee1968stability}\/} (see Figure~\ref{fig:balance_stress} (top));
(ii) a visualization of the stress distribution --- script the model as input to the ANSYS R19.0 (Academic) software (see Figure~\ref{fig:balance_stress} (bottom)); and
(iii) assemblability --- test if all bricks can be iteratively removed from the assembly without collision.
Based on the analysis, we can also generate an assembly sequence that respects the model symmetry, produce scripts to render a model assembly video,
and generate \legomark-style assembly instructions with the help of the LPub3D software~\cite{Sandy-2018-LPub3D}.
\new{See Supplementary material part \final{D} for implementation details.}

\begin{table}[t]
	\centering
	\caption{Statistics of our results:
%
(i) the input sketch complexity shows the total number and total length of sketch lines, and the number of extracted coplanar components (see Section~\ref{ssec:beam_orient}));
%
(ii) statistics of the generated models include the number of beams, total number of bricks (beams \& connecting bricks), and model's physical size; and
%
(iii) our method's running times.}
	\vspace*{-1mm}
	\includegraphics[width=8.3cm]{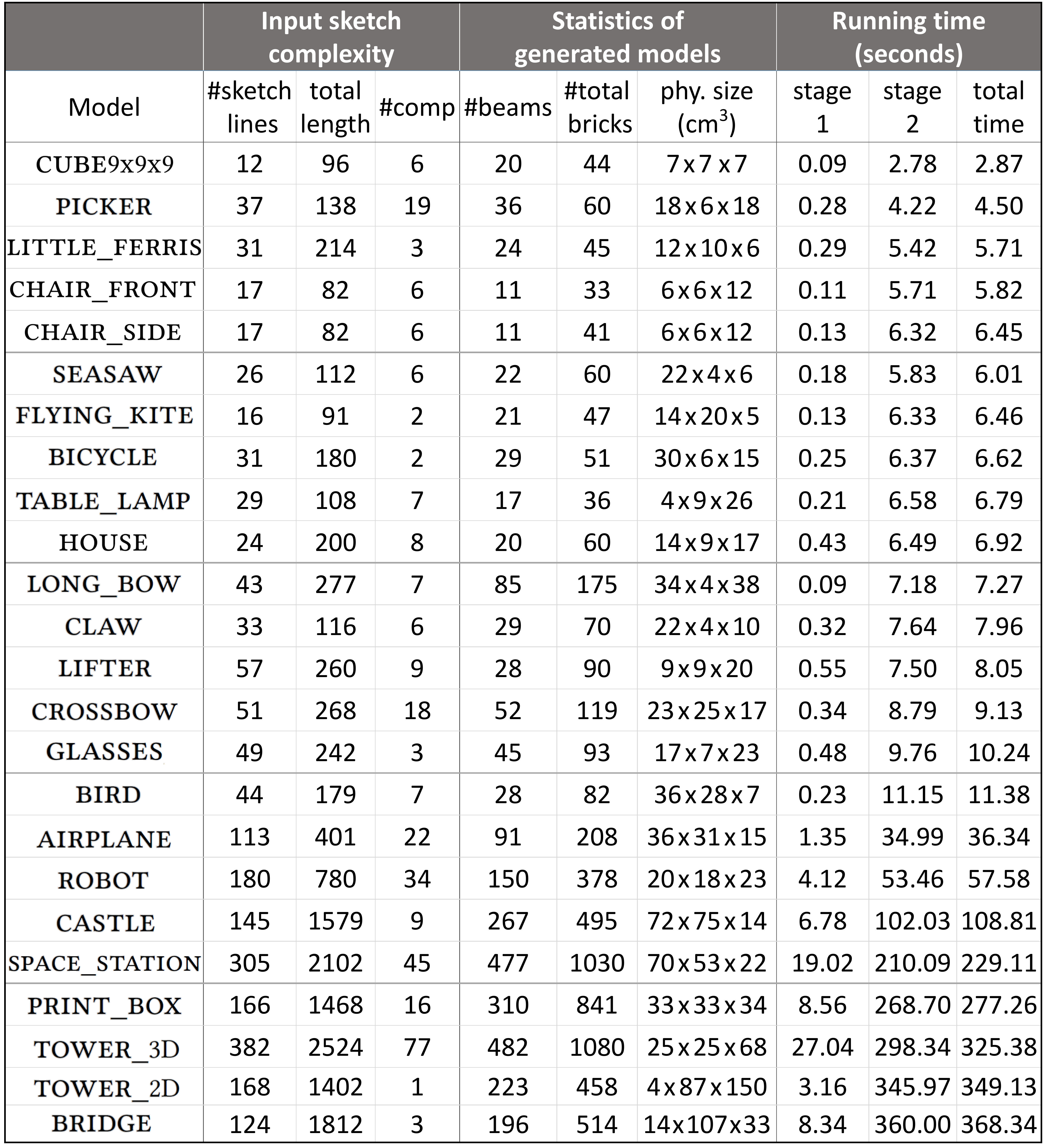}
	\label{fig:statistics}
	\vspace*{-1mm}
\end{table}


\section{Results}
\label{sec:results}
We employed our method to design and generate a rich variety of \legomark Technic models, as listed in Table~\ref{fig:statistics}.
All models are automatically generated from input sketches on a MacBook Pro with a dual-core Intel i5 CPU and 8GB RAM.
%
Figure~\ref{fig:phys_gallery} photographs the physical assemblies of \new{twenty} of the computed models, while Figure~\ref{fig:gallery} shows the renderings of four remaining larger ones.
\new{Nine of them can perform motion dynamics, e.g., {\sc bird}, {\sc claw}, and {\sc long\_bow}.}
%
These results demonstrate that our method is able to generate \legomark Technic models of varying size, shape, structure, \new{and functionality,} from small models, such as {\sc picker}, {\sc flying\_kite} and {\sc lifter}, with less than 100 bricks, to medium-sized models, such as {\sc long\_bow}, {\sc claw}, and {\sc airplane}, as well as to large models with over 400 bricks, such as {\sc castle} and {\sc print\_box}.
Particularly, the results show coherent connections between bricks and the preserved symmetry.
In terms of shape and structure, our method can generate large planar and nearly-planar models like {\sc tower\_{\footnotesize 2D}} and {\sc castle}, as well as 3D structures of varying complexity.
Please refer to the supplementary video for the input sketches and animated results.

\begin{figure*}[!t]
	\centering
	\includegraphics[width=17.6cm]{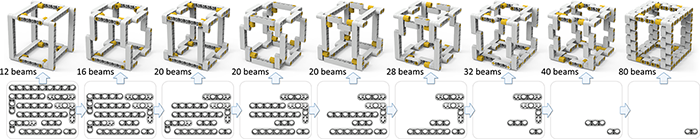}
	\vspace*{-1.5mm}
	\caption{Robustness of our method to brick sets: cubes generated by our method using different brick sets, from full to a single-beam set.}
	\label{fig:brick_set}
	\vspace*{-1mm}
\end{figure*}

\begin{figure}[!t]
	\centering
	\includegraphics[width=8.3cm]{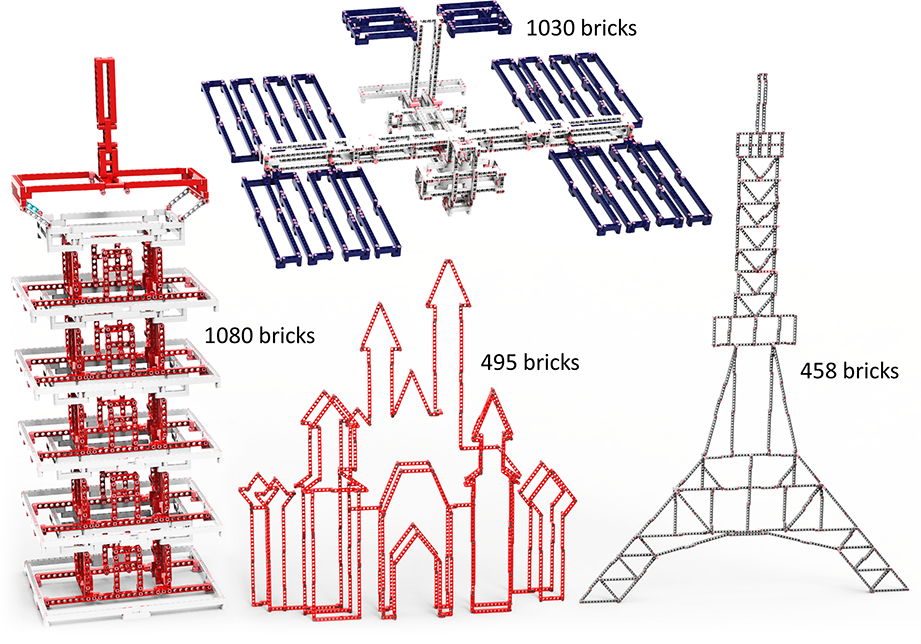}
	\vspace*{-2mm}
	\caption{Larger \legomark Technic models generated by our method.
	\final{From left to right, we have\/} {\sc tower\_{\footnotesize 3D}}, {\sc space\_station\/}, {\sc castle}, and {\sc tower\_{\footnotesize 2D}}.}
	\label{fig:gallery}
\end{figure}

\begin{figure}[!t]
	\centering
	\includegraphics[width=8.35cm]{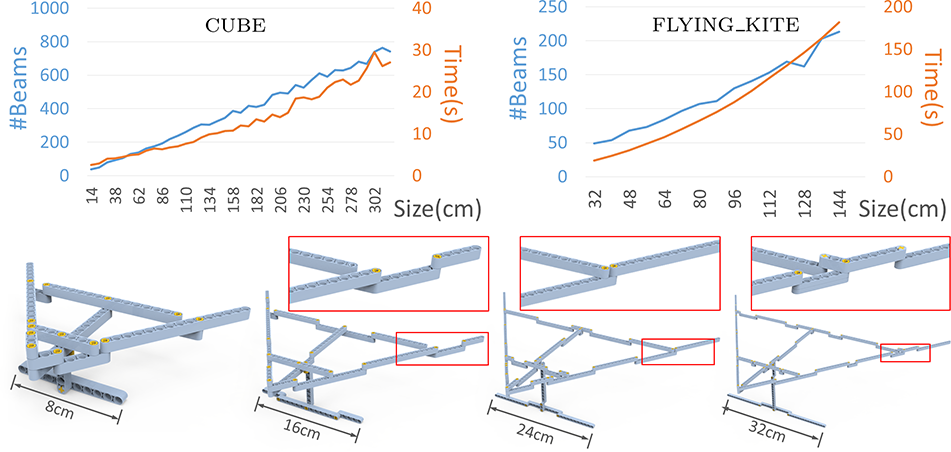}
	\vspace*{-2mm}
	\caption{Scalability test: {\sc cube} and {\sc flying\_kite} in increasing sizes.}
	\label{fig:scalability}
\end{figure}


\vspace*{-3pt}
\paragraph{Timing performance}
The rightmost part in Table~\ref{fig:statistics} reports the running times of our method.
Stage one takes only a few seconds to complete, except for a few very large models like {\sc space\_station\/} and {\sc tower\_{\footnotesize 3D}}.
%
As expected, Stage two takes up most of the processing time, since it needs to iteratively refine the layout and brick connections.
%
Overall, the whole method completes in only a few minutes, even for very large models with a thousand bricks.
Note also that we sort the rows (models) in Table~\ref{fig:statistics} by the total running time to reveal that the running time depends not only on the number of bricks but also on the complexity of the input models.

\vspace*{-3pt}
\paragraph{Scalability}
We examine the scalability of our method by generating models for {\sc cube} and {\sc flying\_kite} in increasing sizes.
Figure~\ref{fig:scalability} shows the statistics of the results, where our method can efficiently generate models in varying scales within minutes, and the number of beams and running time increase roughly linearly.



\vspace*{-3pt}
\paragraph{Robustness to brick set}
To explore our method's robustness to variations in brick set, we start with a full set of beams to generate {\sc cube} with the goal of simple layouts.
Then, we gradually take out the most frequently-used beam from the set and re-generate {\sc cube}, until the set is empty.
From the results shown in Figure~\ref{fig:brick_set}, we can see that our method can produce coherent structures for all different brick sets.
Importantly, beam arrangement is not a simple decision that greedily chooses the longest beam, but a global optimization that considers the overall structural coherency, simplicity, and symmetry.
See the second cube in Figure~\ref{fig:brick_set}, our method can skillfully arrange L-shaped beams to minimize the brick consumption; see Supplementary material part \final{E} for more results.




\vspace*{-3pt}
\paragraph{Efficiency comparison to alternative methods}
To evaluate the efficiency of our method, we compare it with four different methods that are built upon our framework, and use them to generate Technic models, specifically for minimal gap and minimal beam counts:
(i) a {\em random\/} method, which starts with a random layout and applies the layout modification operator to iteratively refine the layout for 50000$n$ times ($n$ is the input problem size to be described later);
%
(ii) a {\em greedy\/} method, which progressively adds locally-optimum beams that cover the most portions of the uncovered sketch;
%
(iii) a {\em beam search\/} method\final{~\cite{medress1977speech}}, which builds a three-layered search tree of partial layouts as internal nodes and iteratively updates the layout with the local best choice using a beam search width of 75; and
(iv) an {\em ant colony\/} optimization method, which generalizes the set cover model in~\cite{ren2010new} to handle beam layering.

In the comparison, the input sketches we employed are 2D uniform grids of $n$-by-$n$ cells ($n$=1..10); each cell is $4$$\times$$4$ sq. units in size.
The difficulty in generating \legomark Technic models for these grids, especially the larger ones, is that we have to minimize {\em both\/} gaps and beam counts, while 
considering beam {\em layering\/}; hence, we cannot simply use the longest beams, which will easily lead to gaps.

\begin{figure}[!t]
	\centering
	\includegraphics[width=8.35cm]{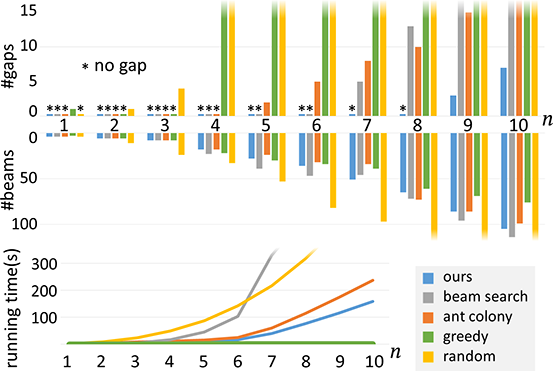}
	\vspace*{-1mm}
	\caption{Compare our method with four alternative methods (see legend) in terms of the number of gaps and beams in use, as well as the running times. The input models are uniform square grids with increasing sizes ($n$).}
	\label{fig:comparison}
\end{figure}

Figure~\ref{fig:comparison} shows the comparison results: the number of gaps and beams, as well as the running times, for different grid sizes ($n$).
Except for greedy and random, most methods can generate gap-free layouts for $n \leq 4$, where simply using the longest beams can effectively produce good solutions.
Interestingly, the random method fails to find gap-free layouts even for $n=2,3$ grids; this reveals the immenseness of the search space.
When $n$ gets larger, random and greedy start to produce layouts with lots of gaps, while ant colony, beam search, and our method can still produce gap-free layouts for $n=4, 6, 8$ grids, respectively.
%
Particularly, ant colony can generate layouts with small number of beams for $n=5,6,7$, but it fails to avoid gaps, since the search space is not only immense but also discrete rather than continuous.
On the other hand, beam search can make good local decisions and avoids gaps better than ant colony; however, it takes much longer time to run and fails to minimize beam counts for large $n$.
In contrast, our method is able to produce gap-free layouts even for large grids ($n$$=$$8$), while effectively minimizing the beam count using much less computing time.

\begin{figure}[t]
	\centering
	\includegraphics[width=8.35cm]{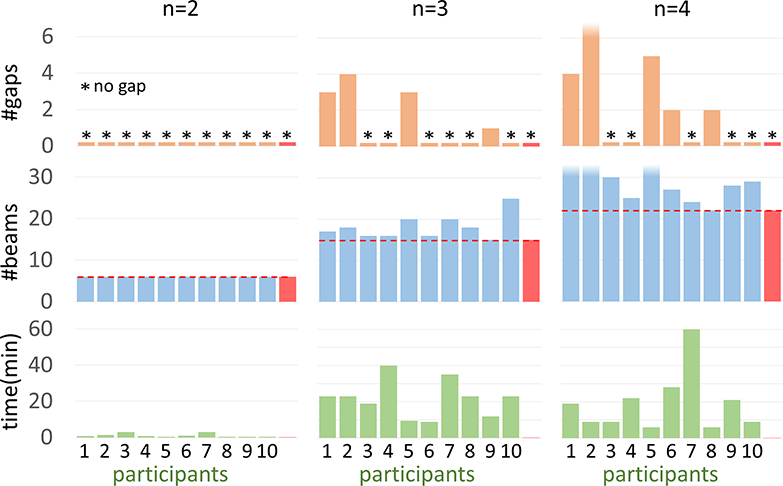}
	\vspace*{-1mm}
	\caption{Comparing manual designs with our method on generating \legomark\\ Technic grids of 2$\times$2, 3$\times$3, and 4$\times$4 cells.
The red bars show the performance of our method, but note that the timing bars are barely visible, since our method was able to compute the solutions in 2.5s, 5.8s, and 8.6s, respectively.}
	\label{fig:human_perform}
\end{figure}

\vspace*{-3pt}
\paragraph{Manual designs}
To obtain a sense of how difficult it is for humans to design \legomark Technic models, we recruited ten participants 
(6 males \& 4 females, aged 22 to 25).
Among them, four had experience in building Technic models.
Here, we followed the above comparison experiment and asked the participants to design Technic models for 2D grids with minimal gaps and beam counts.
However, considering human building effort, we considered only $n=2, 3, 4$ grids, and limited the brick set to contain only beams of up to nine units long.
Before the experiment starts, we printed the grids on A4 papers in the same physical scale as the real bricks, showed the model for $n=1$, and taught the participants the meaning of gaps.
Then, we gave each participant at most an hour to work on each model, and recorded the resulting gap count, beam count, and time spent.

Figure~\ref{fig:human_perform} shows the results; see Supplementary material part \final{F} for the models.
For the smallest grid, all participants could find the optimal solution in a few minutes, while the two larger grids are more challenging: no one was able to find better or even the same solutions as our method, which are gap-free with the minimal beam count.
For the largest grid, $n=4$, only five participants found gap-free solutions, while one of them managed to find a solution with 24 beams, but it was still not as good as the solution produced by our method.
%
Note that before this experiment on 2D grids, we did a pilot study asking the participants to build \legomark Technic models of 3D cubes.
However, those who did not have prior experience failed to connect bricks into cubes; some of them tried it for 50+ minutes.
We concluded that it would be too challenging for non-expert participants to arrange beams in 3D without gaps.



\begin{figure}[!t]
	\centering
	\includegraphics[width=8.35cm]{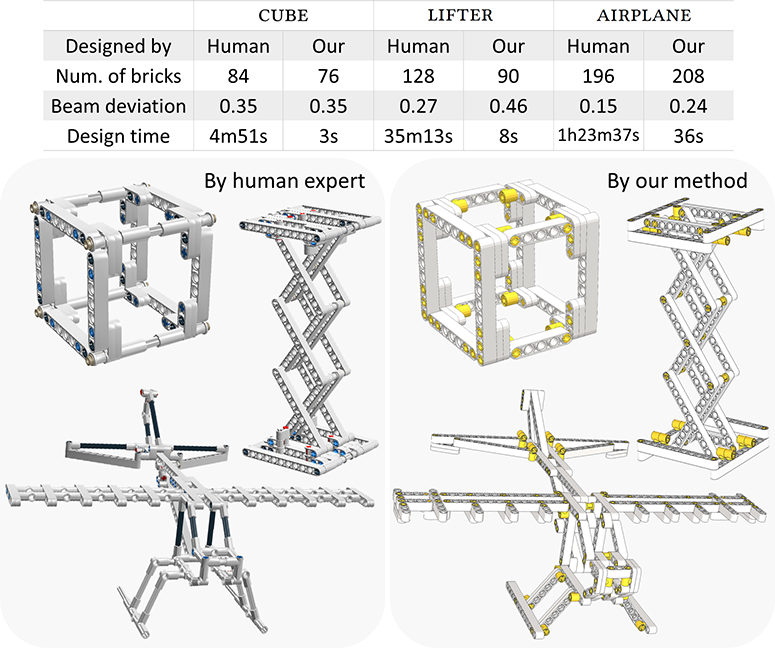}
	\vspace*{-1mm}
	\caption{\new{Models designed by the \legomark Technic expert and by our method.}}
	\label{fig:expert_compare}
\end{figure}


\new{\vspace*{-3pt}
\paragraph{Compare with human expert}
Further, we recruited an expert who had over five-year full-time working experience specialized in \legomark Technic design.
In this experiment, we first showed to him some input sketches without showing him the models generated by our method, then asked him to design models that are faithful to these inputs with the least number of bricks.
His first comment was that such requirement is the same as what he did in his daily designs.
Also, we learnt that he preferred to create his designs using the \legomark Digital Designer software~\shortcite{Lego-2018-Digital-Designer} instead of manual assembly, 
\final{since the software tool provides fast brick retrieval, more brick choices, and can replicate symmetric sub-structures.}
Here, we recorded his design time and the number of bricks in use, and computed the beam deviations in the results.
From the results shown in Figure~\ref{fig:expert_compare}, we can see that the \legomark Technic models produced by our method are similar to those designed by the expert, in terms of beam count, beam deviation, and visual appearance.
However, designing the models using conventional software took the expert minutes to more than an hour, while our method is able to design comparable solutions in less than a minute.\/}

%


\begin{figure}[!t]
	\centering
	\includegraphics[width=8cm]{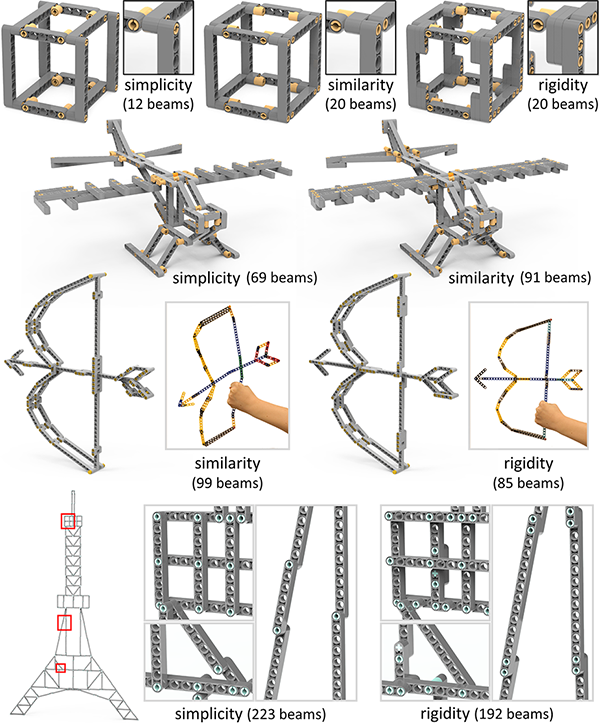}
	\vspace*{-1mm}
	\caption{\legomark Technic models generated with emphasis on different terms in our objective, e.g., simplicity, rigidity, and similarity. Each row shows different results from the same input sketch.}
	\label{fig:preference}
	\vspace*{-2mm}
\end{figure}

%

\vspace*{-3pt}
\paragraph{Designs created by participants using our tool}
We recruited seven participants (5 males \& 2 females, aged 21 to 24) to try our tool.
Among them, two had experience in building \legomark Technic models and one had drawing background.
When they came to the lab, we taught them our GUI and the requirements on the input sketch, then showed them a few example input sketches and the output Technic models.
Then, each participant used around 15 min. to design what they wanted to create, and used around 37 min. on average to sketch their designs on our GUI.
\new{Figure~\ref{fig:phys_gallery} (bottom right)} shows four of the models designed and assembled by the participants, featuring movie characters, everyday objects, and abstract models:
\new{{\sc little\_ferris} (design: 12 min., generation: 6 sec.),
{\sc bicycle} (design: 17 min., generation: 7 sec.),
{\sc glasses} (design: 27 min., generation: 10 sec.), and
{\sc robot} (design: 87 min., generation: 58 sec.).}

\vspace*{-3pt}
\paragraph{Adapting to different objectives}
Our method can generate models from the same input sketch by emphasizing \final{$\mathcal{F}_\text{dev}$, $\mathcal{F}_\text{tbl}$, and $\mathcal{F}_\text{rgd}$ in our objective to different extent\/}; see Section~\ref{ssec:obj} for the specific weight setting.
For example, we can aim for simple layouts with minimized brick count or aim for high input similarity at the expense of using more bricks; see results for {\sc cube} and {\sc airplane} in Figure~\ref{fig:preference}.
Additionally, we can aim for connection rigidity and encourage our method to connect adjacent beams with multiple holes;
see {\sc cube} (rightmost), {\sc long\_bow}, and {\sc tower\_{\footnotesize 2D}} in Figure~\ref{fig:preference}.

\begin{table}[!t]
	\centering
	\caption{\final{Effects of adjusting the weights of objective terms on number of beams ($|\mathcal{B}|$), layer compactness ($\mathcal{F}_\text{cpt}$), model symmetry ($\mathcal{F}_\text{sym}$), average pin-head ratio ($\bar{\rho}$), number of collisions ($N_\text{col}$), number of gaps ($N_\text{gap}$), and number of failure connections ($N_\text{cfail}$).
	We generate each result by independently modifying each weight, $w_\text{cpt}$, $w_\text{sym}$, $w_\text{phr}$, $w_\text{col}$, $w_\text{gap}$, or $w_\text{coh}$, while fixing the others.
	The adjustment either nullifies (null) the effect of the associated term by setting the weight to zero or emphasizes the effect by doubling the weight (x2).
	Compared with the result (last column) generated under the default weights, we highlight the improved aspects in light green and worsened aspects in light pink, showing that deviations from the default settings may improve certain aspects but could worsen others.}}
	\vspace*{-2mm}
	\includegraphics[width=8.1cm]{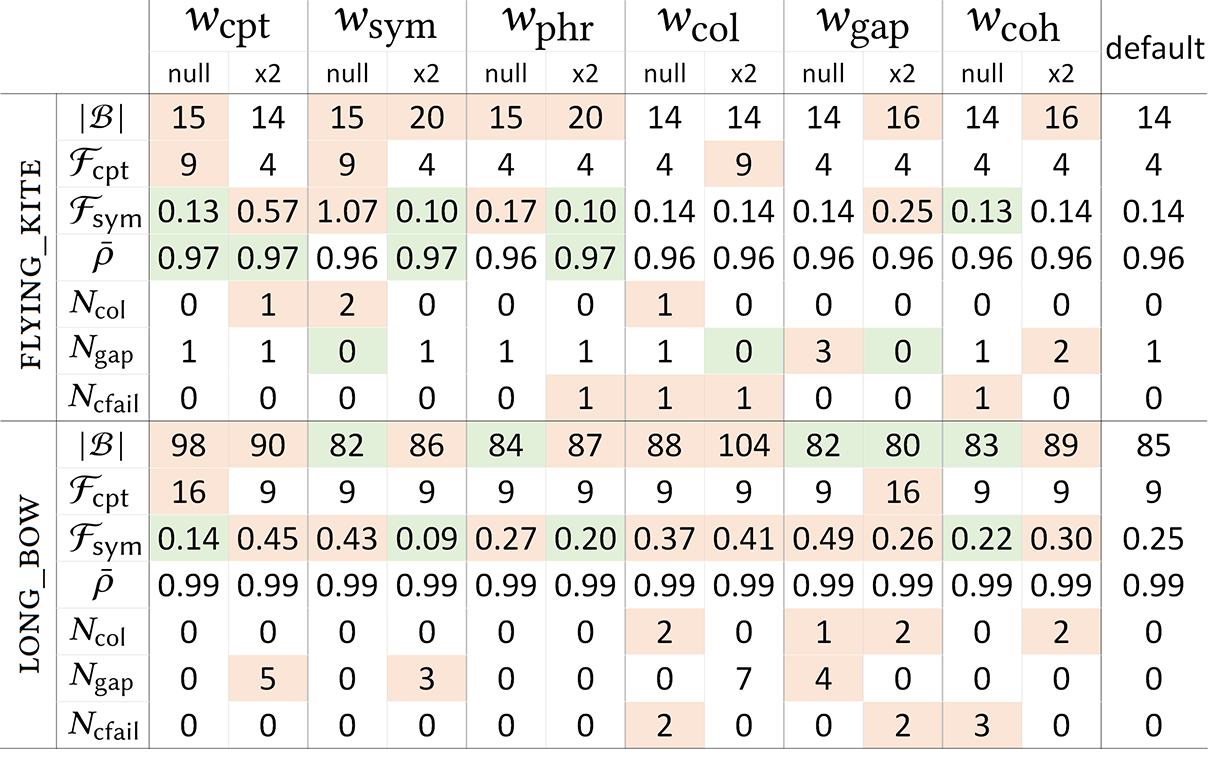}
	\label{fig:abla_study}
	\vspace*{-1mm}
\end{table}

\begin{figure}[!t]
	\centering
	\vspace*{-2mm}
	\includegraphics[width=8.35cm]{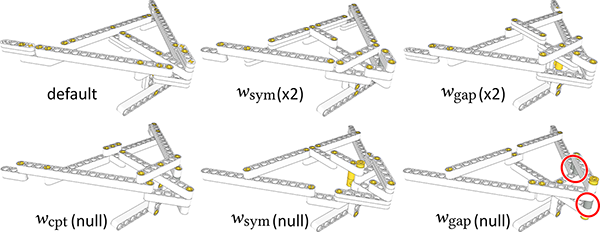}
	\vspace*{-2mm}
	\caption{\final{Some of the generated models for {\sc flying\_kite} in Table~\ref{fig:abla_study}.}}
	\label{fig:vis_abla_study}
	\vspace*{-3mm}
\end{figure}

\begin{figure*}[!t]
	\begin{minipage}[c]{0.635\textwidth}
		\setcaptionwidth{1.0\textwidth} 
		\centering
		\includegraphics[width=0.99\textwidth]{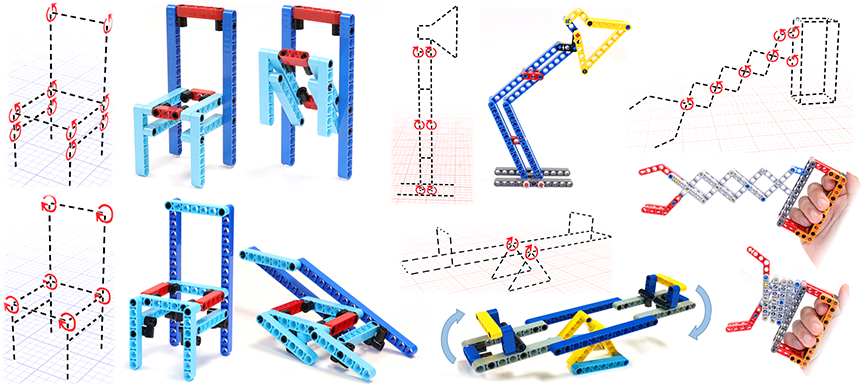}
		\vspace*{-1.5mm}
		\caption{\new{Example models with annotated joint rotations generated from sketches.
		From left to right, we have {\sc chair\_front} (top left), {\sc chair\_side} (bottom left), {\sc table\_lamp}, {\sc seasaw}, and {\sc claw}.} \legoclaim}
		\label{fig:dynamic_models}
	\end{minipage}%
	\hspace{3mm}
	\begin{minipage}[c]{0.325\textwidth}
		\setcaptionwidth{1.0\textwidth}
		\centering
		\includegraphics[width=0.99\textwidth]{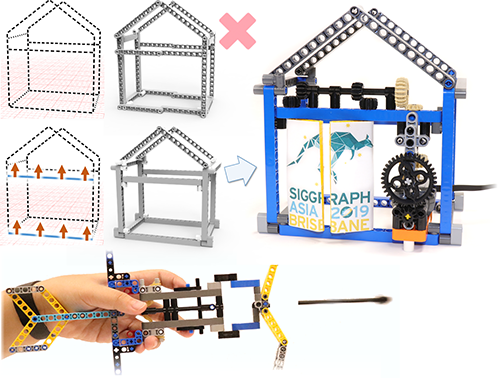}
		\vspace*{-7mm}
		\caption{\new{Top: {\sc house} directly generated from the input sketch.
Middle: beams are constrained to exactly pass through the annotated sketch line (in blue) for embedding a gear system.
Bottom: {\sc long\_bow} embedded with a shooting mechanic.} \legoclaim}
		\label{fig:embed_dynamic}
	\end{minipage}
	\vspace*{-2mm}
\end{figure*}

\vspace*{-3pt}
\paragraph{\rzf{Ablation study on objective terms}}
\final{Further, we conduct an experiment to show how other terms in the objective function affect the results.
Here, we fix $\{ w_\text{dev}, w_\text{tbl}, w_\text{rgd} \}$ as $\{0,100,0\}$ to ensure model simplicity, then test each of the other objective terms by independently adjusting its associated weight. We set the weight to zero to assess the necessity of an objective term, or double its value to increase its impact.
%
Table~\ref{fig:abla_study} summarizes the ablation study results on two input models.
Without changing the annealing temperatures and cooling rate, we found that doubling any particular weight usually breaks the balance among the objective terms and could deteriorate the results.
On the other hand, neglecting a particular term could deteriorate the corresponding aspect in the resulting model, without improving much on the other aspects.
Figure~\ref{fig:vis_abla_study} shows some of the models generated in this ablation study.}

%

\vspace*{-3pt}
\phil{\paragraph{Dynamic models}
Figure~\ref{fig:dynamic_models} shows physical assemblies designed with annotated joint rotations.
Comparing the two {\sc chairs\/}, we can see that by annotating joints with different rotation axes, our method can constrain the connecting beams to rotate at specific orientations at the joints, thus leading to the production of different models.
%
%
Besides, given a mechanical system (see the two examples in Figure~\ref{fig:embed_dynamic}), we can force the generated beams in the layout to exactly pass through some annotated sketch lines, such that we can then embed a mechanical system into the generated model.
Please see the supplementary video for these models in action.\/}


\new{\vspace*{-3pt}}
\paragraph{Discussion on global rigidity}
If we optimize the \final{{\sc long\_bow}} model without the rigidity term, the model could be deformed due to the gravity, as demonstrated in Figure~\ref{fig:preference} (middle left).
To evaluate the global rigidity of a \legomark Technic structure is a very challenging problem.
First, global non-rigidity is a result of multiple (a subset of) joints in the overall structure.
Particularly, a joint may be transitively (or indirectly) blocked to rotate by others that are not located next to it.
Here, trying every joint subset would require tremendous computation.
Also, we have to deal with a diverse and irregular brick set that can be connected in many different ways.
Further, we need a unified model to evaluate the effectiveness of individual connections, brick-blocking relations, and other physical constraints such as friction.
Hence, we believe analytically evaluating the global rigidity is very challenging, and will require extensive works.

In the course of this work, we have thought of two approaches to this problem:
(i) apply external forces on the structure, then see if the structure deforms in a physical simulation; and
%
(ii) formulate a motion-based equation by setting velocity variables on every joint, constrain them based on the beam connections, solve it, and locate the joints with non-zero velocities.
%
Clearly, these approaches are preliminary and require more thoughtful ideas to turn them into solid feasible solutions.
Hence, we only consider local rigidity in this work, and leave global rigidity as our future work.
\section{Conclusion, limitation, and future work}

We presented a first attempt to computerize \legomark Technic constructions.
Altogether, there are three contributions in this work.
First is an automatic computational method that can efficiently generate \legomark Technic models from user input sketches.
Particularly, the generated model is a coherently-connected structure composed of \legomark Technic bricks, and we can aim for faithfulness to input sketch, model simplicity, and structural integrity in the model generation.
%
Second is our comprehensive model for various aspects in \legomark Technic constructions, including the enumeration of brick properties and connection mechanisms, conceptualization of the input sketch as a guiding graph, formulation of the construction requirements into an objective, \phil{and dynamic model constructions with hinge-style rotations and dynamic parts embedding\/}.
Third, we also developed a working system to sketch the inputs, and to analyze the balance, stress, and assemblability of the generated model.
%
In the end, we employed our system to create \legomark Technic constructions of various shapes, complexities \phil{and functionalities\/}, compared it with four alternative methods, \phil{general users and a human expert\/}, evaluated it for scalability, robustness and adaptiveness, as well as physically built most of the generated models.


\begin{figure}[!t]
	\centering
	\includegraphics[width=8.2cm]{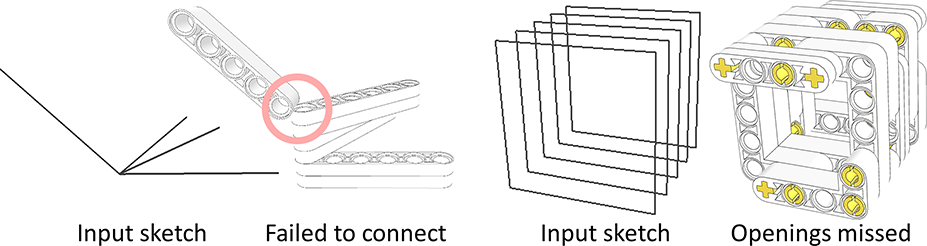}
	\vspace*{-2mm}
	\caption{Failure cases from our current computational method: \new{intersecting sketch lines\/} (left) and closely-packed sketch lines (right).}
	\label{fig:failure}
	\vspace*{-2mm}
\end{figure}

\vspace*{-3pt}
\paragraph{Limitations}
\final{In terms of model generation,} \rzf{while} our method makes an effort to adjust the beams,
it may still fail to create connections in some situations, especially when several beams intersect/touch one another non-orthogonally; see Figure~\ref{fig:failure} (left).
For dense and parallel sketch lines, the generated models may not retain the gaps between the lines; see Figure~\ref{fig:failure} (right).
\final{Also, our current method assumes most sketch lines are covered by the beam bricks. However, as shown in the expert's designs, axles may also be used to cover the sketch lines, where some special axle-related connector bricks can be used in the connections.
\rzf{Moreover,} as discussed earlier, our current formulation for connection rigidity is local, not global.
\final{Lastly, in our sketching tool, diagonal lines in the sketches should follow certain Pythagorean triples, e.g., after we sketch an L-structure with two orthogonal lines, if we want to use an extra line to diagonally connect the previous two lines, the length of the diagonal line may be five-unit long, since $5^2=3^2+4^2$.}

\rzf{In terms of our handling of dynamic constraints in the input, while} our method realizes user-annotated hinge-style rotations, the decision of where to put joints and how to make different parts work together to realize a desirable dynamic behavior still remains hard for novice users. 
\rzf{Also,} besides hinge-style rotations, other \legomark Technic motions such as sliding, sheering, lifting, and their combinations are not yet considered in our system.\/}

\vspace*{-3pt}
\paragraph{Discussion and future work}
\final{Addressing the \rzf{various limitations above} already suggests a comprehensive \legomark Technic design system involving a number of sub-problems, such as global rigidity analysis, rigid structure generation, inverse joints computation, and inverse multi-functionality design with mechanical elements.}
%
Moreover, official \legomark Technic models often contain customized parts specially-designed for the outer shell of the model. We would also like to design and fabricate customized 3D-printed parts to work with the beams and connectors in the \legomark Technic system for driving 3D-printed customized models.

\begin{acks}
	\final{We thank all the anonymous reviewers for their comments and feedback. We also acknowledge help from Ruihui Li for UI development, Shufang Wang and Tianwen Fu for model assembly, Chun Yu Liu for designing the models shown in Figure~\ref{fig:expert_compare}, and Wallace Lira and Johannes Merz for paper proofreading.
		Figures 2 (a), (b) and (c) are courtesy of YouTuber TECHNICally Possible, Nicolas Lespour, and Will Gorman, respectively.
		This work is supported in part by grants from the Research Grants Council of the Hong Kong Special Administrative Region (Project no. CUHK 14201918 and 14203416), NSERC grants (No. 611370), and gift funds from Adobe.
	}
\end{acks}


\bibliographystyle{ACM-Reference-Format}
\bibliography{comptechnic}

\end{document}